\documentclass[12pt]{article}
\usepackage{multirow}
\usepackage{amsmath,amsfonts,amssymb,bm,slashed,bbm,mathrsfs}
\usepackage{graphicx,color}
\renewcommand{\Ref}[1]{(\ref{#1})}
\newcommand{\eq}[2]{\begin{align}\label{#1}#2\end{align}}

\newcommand{\nn}{\nonumber}
\renewcommand{\ni}{\noindent}
\newcommand{\pa}{\partial}

\newcommand{\om}{\omega}

\newcommand{\sig}{\sigma}
\newcommand{\Ga}{\Gamma}

\begin{document}

\title{Effective potential of gluodynamics in background of Polyakov loop and colormagnetic field}
\author{M. Bordag\thanks{bordag@uni-leipzig.de}\\
{\small Institute for Theoretical Physics, Universit{\"a}t Leipzig}\\[9pt]
 V. Skalozub\thanks{e-mail: Skalozubv@daad-alumni.de}\\
{\small Oles Honchar Dnipro National University, 49010 Dnipro, Ukraine}}

\date{December 2, 2021}
\maketitle\thispagestyle{empty}
%
\begin{abstract}
In SU(N) gluodynamics,   above the  de-confinement temperature, the effective potential has minima at non-zero $A_0$-background fields in the two-loop approximation. Also, it has a minimum at non-zero chromomagnetic background field, known as 'Savvidy'-vacuum, which shows up on the one-loop level. In this paper, we join these two approaches. We formulate, at finite temperature,  the effective action, or the free energy,  in SU(2) gluodynamics on the  two-loop level, with both, $A_0$ background and magnetic background present at the same time, which was not done so far. We provide the necessary representations for both, effective numerical calculation and high-temperature expansions. 

The results are represented as a 3D plot of the real part of the effective potential. Also, we reproduce for zero either, the $A_0$-background or the magnetic background, the known minima and compare them.
The imaginary part is, on the two-loop level, still present. We mention that, as is known from literature for the case without $A_0$-background,  the imaginary part is compensated by the ring ('daisy') diagrams. However, our results reveal an unnatural, singular behavior of the effective potential in the region, where the imaginary part sets in.  Our conclusion is that one has to go beyond the two-loop approximation  and its ring improved version, in order to investigate the minimum of the effective action as a function of  $A_0$ and chromomagnetic field, and its stability, at least in the approximation of superdaisy diagrams, i.e., the Hartree approximation in the CJT formalism.

\end{abstract}

\section{\label{T1}Introduction}
Background fields and classical solutions in QCD are an important topic towards a theory of confinement. There is quite a number of them which were investigated as, for example, instantons, monopoles or vortices  \cite{gros81-53-43}. A special role plays constant background fields, to a large extent since these are frequently solutions of the classical field equations and allow in many cases for quite explicit formulas.
One prominent example is the $A_0$-background, which is closely related to the Polyakov loop. At finite temperature, such field cannot be gauged away and was intensively investigated beginning with \cite{weis82-25-2667}. In the early 90-ies, two-loop contributions were calculated and with these, as was shown in \cite{skal94-57-324}, the effective potential has non-trivial minimums and related condensate fields.  These form a hexagonal structure in the plane of the relevant components $A_0^3$ and $A_0^8$ of the background field. These condensate fields constitute the so-called '$A_0$-vacuum'. This topic is also intensively investigated by functional approaches, for a recent work see \cite{gaof21-103-094013} and citations therein.

A different kind of background field is the chromomagnetic one. Its history dates back to the work \cite{savv77-71-133}, where a minimum of the effective potential in the background of a chromomagnetic field was found, also forming a condensate. This state is called 'Savvidy vacuum'. However, quite soon, in \cite{niel78-144-376} and \cite{skal78-28-113},  it was found to be unstable due to the tachyonic mode. By now, there is a large number of attempts towards stabilization. We mention here only the re-summation of ring ('daisy') diagrams in \cite{skal00-576-430}, resulting in the statement that the imaginary part disappears and the minimum of the real part remains.

There are only  very few attempts to consider $A_0$- and magnetic backgrounds together, \cite{star94-322-403} and \cite{meis02-66-105006}. Both are on the one-loop level. It must be mentioned that this is clearly insufficient as the $A_0$ condensate starts to appear only on the two-loop level.

There are lattice calculations with both backgrounds.  In \cite{demc13-21-13} it was observed that in the presence of a constant color magnetic field the Polyakov loop acquires  a non-trivial spatial structure    along the   direction of the magnetic field. More interestingly, in  \cite{demc08-41-165051}  a common spontaneous generation of both fields was found.

As a step towards the simultaneous generation of both background fields in a perturbative approach, we consider both these background fields on the two-loop level. More specifically, we calculate the effective potential as a function of both parameters, $A_0$, and $B$, in $SU(2)$ gluodynamics. The general expressions for the effective potential,   are generalized  to include the magnetic background in two-loop order. The mathematical tools for the calculation  of the appearing expressions, which can be found to a large extent in the literature, will be represented in a coherent form. This way, we are able to calculate  numerically the effective potential in the $(A_0,B)$-plane for finite   temperature.
Also, we consider the limiting cases $A_0=0$ and $B=0$ in detail and find, for instance, the magnetic condensate in two-loop order, which was also considered in  \cite{skal00-576-430}.
We mention, that a spontaneous generation  of a background field is  meant in the sense, that for the corresponding field the effective action has a minimum below zero, which is energetically favorable. Thereby it is assumed that no external sources are present.

In our investigation, we are mainly interested in the minimum of the effective potential, i.e., on a spontaneous generation, on the two-loop level. At this place, we do not touch the problem of the stability. However, we will observe that the compensation of the imaginary part, found in \cite{skal00-576-430} in an one-loop approximation, is insufficient in the two-loop level.

In  the gluonic sector, in the case of SU(2), the effective potential reads
\eq{W2}{ W^{SU(2)}_{gl} &=B_4(0,0)+2B_4\left(a,b\right)
\\\nn&~~~	+2{g^2}\left[
	B_2\left(a,b\right)^2
	+2 B_2\left(0,b\right) B_2\left(a,b\right)	\right]
	-4{g^2}(1-\xi) B_3\left(a,b\right)B_1\left(a,b\right)
}
with the notation
\eq{ab}{a=\frac{x}{2}=\frac{g A_0}{2\pi T},~~~b=gH_3.
}
Since we work at finite temperature, $ W^{SU(2)}_{gl}$ is, of course, equivalent to the free energy.
The functions $B_n(a,b)$ are defined by
\eq{3}{ B_4(a,b) &= T\sum_\ell\int\frac{dk_3}{2\pi}\frac{b}{4\pi}\sum_{n,\sig}
	\ln\left(\left(2\pi T(\ell+a)\right)^2+k_3^2+b(2n+1+\sig-i0)\right),
\\\nn
B_3(a,b) &=
T\sum_\ell\int\frac{dk_3}{2\pi}\frac{b}{4\pi}\sum_{n,\sig}
\frac{\ell+a}{\left(2\pi T(\ell+a)\right)^2+k_3^2+b(2n+1+\sig-i0)}
\\\nn
	 B_2(a,b) &= T\sum_\ell\int\frac{dk_3}{2\pi}\frac{b}{4\pi}\sum_{n,\sig}
	\frac{1}{\left(2\pi T(\ell+a)\right)^2+k_3^2+b(2n+1+\sig-i0)},
\\\nn		
	B_1(a,b) &=
	T\sum_\ell\int\frac{dk_3}{2\pi}\frac{b}{4\pi}\sum_{n,\sig}
	\frac{\ell+a}{\left(\left(2\pi T(\ell+a)\right)^2+k_3^2+b(2n+1+\sig-i0)\right)^2}.
}
In these formulas the summations run  $n=0,1,\dots$, $\sig=\pm2$ and $\ell$ runs over all integers. The $'-i0'$-prescription  defines the sign of the imaginary part for the tachyonic mode.
These formulas and eq. \Ref{W2} are the generalization of the corresponding two-loop expressions in \cite{enqv90-47-291}, eqs. (3.8) and (A.2)-(A.5), \cite{bely91-254-153}, eq. (14),  \cite{skal92-7-2895}, eq. (4),  and also \cite{skal21-18-738}, eq. (4),
to the inclusion of the magnetic field. Note a "-" sign in \Ref{1.1}). Below we will use also the relations
\eq{3a}{B_3(a,b) &= \frac{1}{4\pi T}\pa_a B_4(a,b),
	~~~&	B_1(a,b) &= \frac{-1}{4\pi T}\pa_a B_2(a,b),
}
which are quite convenient.

For $b\to0$ we note $\frac{b}{4\pi}\sum_{n,\sig}\to\int\frac{d^2k}{(2\pi)^2}$ and get at $b=0$
\eq{1.1}{ 	B_4(a,0)  = \frac{2\pi^2T^4}{3} B_4(a),
	~~ 		B_3(a,0)  =\frac{2\pi T^3}{3}	B_3(a),
 ~~~  		B_2(a,0)  = \frac{T^2}{2}		B_2(a),
 ~~~		B_1(a,0)  =-\frac{ T}{4\pi}	     B_1(a),
}
where $B_n(a)$ are the Bernoulli polynomials, periodically continued.
The special values for, in addition, $a=0$ are
\eq{4}{ B_4(0,0)  = -\frac{\pi^2T^4}{45},
	~~~   B_3(0,0)= 0,
	~~~   B_2(0,0)= \frac{T^2}{12},
	~~~   B_1(0,0)= \frac{T}{8\pi}.
}
We note that these formulas hold for $T>0$. For the $T\!=\!0$--case see section \ref{T2.1} together with the renormalization.
The motivation for the above choice of the notations  is that  the functions $B_n(a,b)$, \Ref{3}, are  the corresponding mode sums without additional factors.

The rest of the paper is organized as follows. In the next section we give, in detail, representations of the functions $B_j(a,b)$, \Ref{3}, including their high temperature expansions. In the third section, we investigate the minima of the effective action in the pure magnetic case and, in Section \ref{T4} we consider the pure $A_0$-case and compare both. Section \ref{T5} represents with the effective potential as a function of both, $a$ and $b$ the main result of the paper, which is discussed in the last section.

\ni Throughout the paper we use natural units with $\hbar=c=k_{\rm B}=1$.

\section{\label{T2}Representation of the basic functions}
In this section, we consider in detail the functions defined in \Ref{3}.
First of all, we write down their  proper time representations,
\eq{2.1}{
B_4(a,b) &=-\pa_s \int_0^\infty\frac{dt}{t}\frac{t^{s-\frac12}}{\sqrt{4\pi}\Ga(s)}
\frac{b}{4\pi}\sum_{n,\sig}	T\sum_\ell \exp\left\{-t\left[(2\pi T(\ell+a))^2+b(2n+1+\sig-i0)\right]\right\},
\\\nn
 B_2(a,b) &= \int_0^\infty\frac{dt}{t}\frac{t^{s+\frac12}}{\sqrt{4\pi}\Ga(s+1)}
\frac{b}{4\pi}\sum_{n,\sig}	T\sum_\ell \exp\left\{-t\left[(2\pi T(\ell+a))^2+b(2n+1+\sig-i0)\right]\right\}.
}
Here, $s$ is the regularization parameter with $s\to 0$ at the end. The '-i0'-prescription is for the tachyonic mode. The arbitrary parameter $\mu$ which comes in with dimensional regularization is not shown in these formulas.
We assume for the moment $\sig>-1$ having in mind the analytic continuation to $\sig=-2$ later. In \Ref{2.1} we have already integrated over $k_3$. The next step is the application of the Poisson resummation formula which results in
\eq{2.2}{
B_4(a,b) &=-\pa_s \int_0^\infty\frac{dt}{t}\frac{t^{s-1}}{{4\pi}\Ga(s)}
\frac{b}{4\pi}\sum_{n,\sig}	
\sum_N \cos(2\pi  aN)\exp\left\{-\frac{N^2}{4T^2t}-tb(2n+1+\sig-i0) \right\},
\\\nn
B_2(a,b) &= \int_0^\infty\frac{dt}{t}\frac{t^{s}}{{4\pi}\Ga(s+1)}
\frac{b}{4\pi}\sum_{n,\sig}	
\sum_N \cos(2\pi aN)\exp\left\{-\frac{N^2}{4T^2t}-tb(2n+1+\sig-i0) \right\}.
}
The summation over $N$ runs over all integers.

These expressions allow for several ways to proceed. However, before doing this, one needs to split into $T=0$ -- parts and temperature-dependent part. Afterward, one may choose a representation in terms of Bessel functions, or in terms of Theta functions. In addition, before doing so we must do another split. This is, as we will see below, necessary because of the exceptional role of the tachyonic mode which requires a separate treatment.
So we define
\eq{2.3}{B_x(a,b)&= B_x^{T=0}(a,b)+\Delta_T B_x(a,b),
}
where $x=1,2,3,4$ denotes the functions in \Ref{3}, as split into $T=0$  and temperature dependent parts. The latter we split  further,
\eq{2.4}{\Delta_T B_x(a,b) &=\Delta_T B_x^{nt}(a,b)+\Delta_T B^{ta}_x(a,b),
}
where $\Delta_TB^{nt}_x(a,b)$ denotes the contribution from the non-tachyonic modes, ($(\sig=+2,~n=0,1,\dots)$ and $(\sig=2,~n=1,2,\dots)$), and $\Delta_T B^{ta}_x(a,b)$ denotes the contribution from the tachyonic mode, ($\sig=-2,~n=0$).

In the next subsections, we consider these contributions separately.

\subsection{\label{T2.1}The zero temperature part}
In eq. \Ref{2.2}, the zero temperature part results from the $N=0$-term. Carrying out the summations over $n$ and $\sig$ using
\eq{2.1.0}{ \sum_{n,\sig}	e^{-tb(2n+1+\sig)} &=
	e^{tb}+e^{-tb}\coth(tb),
}
where the first term in the right side results from the tachyonic mode,
we arrive at
\eq{2.1.3}{B_4^{T=0}(a,b) &=-\pa_s \int_0^\infty\frac{dt}{t}\frac{t^{s-1}}{{4\pi}\Ga(s)}
	\frac{b}{4\pi} \left(e^{tb}+e^{-tb}\coth(tb)\right),
\\\nn
B_2^{T=0}(a,b) &=  \int_0^\infty\frac{dt}{t}\frac{t^{s}}{{4\pi}\Ga(s+1)}
\frac{b}{4\pi} \left(e^{tb}+e^{-tb}\coth(tb)\right).
}
The first term in the parentheses results from the tachyonic mode and it must be understood in the sense of an analytic continuation. This can be accounted for by the substitution $t\to t e^{i\pi}$ and we get
\eq{2.1.4}{B_4^{T=0}(a,b) &=-\pa_s \int_0^\infty\frac{dt}{t}\frac{t^{s-1}}{{4\pi}\Ga(s)}
	\frac{b}{4\pi} e^{-tb}\left(e^{i\pi(s-1)}+\coth(tb)\right),
	\\\nn
	B_2^{T=0}(a,b) &=  \int_0^\infty\frac{dt}{t}\frac{t^{s}}{{4\pi}\Ga(s+1)}
	\frac{b}{4\pi}e^{-tb} \left(e^{i\pi(s-1)}+\coth(tb)\right).
}
These integrations can be carried out,
\eq{2.1.5}{B_4^{T=0}(a,b) &=-\pa_s {b^{2-s}\mu^{2s}}
	\frac{1+e^{i\pi s}+2(2^{1-s}-1)\zeta(s-1)}{(4\pi)^2(1-s)},
\\\nn
B_2^{T=0}(a,b) &={b^{1-s}\mu^{2s}}
	\frac{-1+e^{i\pi s}-2(2^{-s}-1)\zeta(s)}{(4\pi)^2 s},
}
where we introduced also the arbitrary dimensional parameter $\mu$ which sets the scale. These terms contain  ultraviolet divergences. In the considered case, these can be removed by performing the limit $s\to 0$. We get
\eq{2.1.6}{B_4^{T=0}(a,b) &= -\frac{ib^2}{16\pi}
	+\frac{11b^2}{96\pi^2}\left(\ln\frac{b}{\mu^2} -1 \right),
	\\\nn
B_2^{T=0}(a,b) &=\frac{ib}{16\pi}-\frac{b\ln 2}{16\pi^2},
\\[4pt]\nn
B_1^{T=0}(a,b) &=0,~~~~B_3^{T=0}(a,b) =0.
}
In the first line, we redefined $\mu$ for later convenience. In the following, especially in the pictures, we put $\mu=1$ and assume all quantities to be in arbitrary units.

\subsection{\label{T2.2}Representation in terms of Bessel functions and asymptotic expansion for high $T$}
We continue with the temperature dependent parts.
The representation in terms of Bessel functions appears after the substitution
$t\to t\frac{N}{2T\sqrt{b(2n+1+\sig)}}$, in   \Ref{2.2}, and application of formula 8.4327 (with $z=1$) from \cite{grad07},
\eq{2.8}{\Delta_T B_4(a,b) &= -\frac{b^{\frac32}T}{2\pi^2}
	\sum_{N=1}^\infty \sum_{n,\sig}	\cos(2\pi  aN)
	\frac{\sqrt{2n+1+\sig}}{N}K_1\left(\frac{\sqrt{b}}{T}N\sqrt{2n+1+\sig}\right)
,
\\\nn
\Delta_T B_3(a,b) &= \frac{b^{\frac32}}{4\pi^2}
\sum_{N=1}^\infty \sum_{n,\sig}\sin(2\pi  aN)
\frac{\sqrt{2n+1+\sig}}{N}K_1\left(\frac{\sqrt{b}}{T}N\sqrt{2n+1+\sig}\right)
,
\\\nn \Delta_T B_2(a,b) &= \frac{b }{4\pi^2}
\sum_{N=1}^\infty \sum_{n,\sig}\cos(2\pi  aN)
K_0\left(\frac{\sqrt{b}}{T}N\sqrt{2n+1+\sig}\right)
,
\\\nn
\Delta_T B_1(a,b) &= \frac{b T^{-1}}{8\pi^2}
\sum_{N=1}^\infty \sum_{n,\sig}N\sin(2\pi  aN)
 {\sqrt{2n+1+\sig}} K_0\left(\frac{\sqrt{b}}{T}N\sqrt{2n+1+\sig}\right),
}
where $K_\nu(z)$ are modified Bessel functions. These formulas provide a representation as fast converging double sums. The convergence comes from the decrease of the Bessel function for the non-tachyonic modes. The asymptotics for small $T$, or equivalently, large $b$, has only exponentially small contributions. However, in the other direction the convergence slows down and this representation becomes ineffective. The contribution from the tachyonic mode can be done in the above-mentioned way and the fast decreasing modified Bessel functions turn into Hankel functions. These have an oscillating behavior which makes the representation \Ref{2.8} less effective for them. This is one reason for the above-mentioned splitting of our basic function.  Below, we consider different representations for the tachyonic modes, including the representation in terms of Hankel functions (Sect. \ref{T2.4.3}).

In order to get a high-T expansion from \Ref{2.8} one may apply the Mellin representation of the Bessel functions
\eq{Me}{ K_\nu(z)&=\frac14\int\frac{ds}{2\pi i}
	\ \Gamma\left(\frac{s+\nu}{2}\right)\Gamma\left(\frac{s-\nu}{2}\right)
	\left(\frac{z}{2}\right)^{-s},
}
where the integration path goes parallel to the imaginary axis to the right of all poles of the integrand. Using this representation in \Ref{2.8} and exchanging the orders of integration and summation, we get
\eq{2.9}{\Delta_T B_4(a,b) &= -\frac{b^{\frac32}T}{8\pi^2}
	\int\frac{ds}{2\pi i} \ \Gamma\left(\frac{s+1}{2}\right)\Gamma\left(\frac{s-1}{2}\right)
	\left(\frac{2T}{\sqrt{b}}\right)^s
	\Sigma_N(s)\Sigma_b(s),
\\\nn
\Delta_T B_3(a,b) &= \frac{b^{\frac32}}{16\pi^2}
\int\frac{ds}{2\pi i} \ \Gamma\left(\frac{s+1}{2}\right)\Gamma\left(\frac{s-1}{2}\right)
\left(\frac{2T}{\sqrt{b}}\right)^s
\Sigma^a_N(s)\Sigma_b(s),
\\\nn   \Delta_T B_2(a,b) &= \frac{b}{16\pi^2}
\int\frac{ds}{2\pi i} \ \left(\Gamma\left(\frac{s}{2}\right)\right)^2
	\left(\frac{2T}{\sqrt{b}}\right)^s
\Sigma_N(s-1)\Sigma_b(s+1),
\\\nn   \Delta_T B_1(a,b) &=- \frac{bT^{-1}}{32\pi^2}
\int\frac{ds}{2\pi i} \ \left(\Gamma\left(\frac{s}{2}\right)\right)^2
\left(\frac{2T}{\sqrt{b}}\right)^s
\Sigma^a_N(s-1)\Sigma_b(s+1),
}
where we defined
\eq{2.10}{ 	\Sigma_N(s) &= \sum_{N=1}^\infty\frac{\cos(2\pi  a N)}{N^{s+1}}
	=\frac12
	\left( {\rm Li}_{s+1}\left(e^{ 2\pi i  a}\right)
	+{\rm Li}_{s+1}\left(e^{- 2\pi i  a}\right)
	\right),
&	\Sigma^a_N(s) &= \sum_{N=1}^\infty\frac{\sin(2\pi  a N)}{N^{s+1}} ,
\\\nn
		\Sigma_b(s) &= \sum_{n,\sig}\left(2n+1+\sig\right)^{\frac{1-s}{2}}
		=i^{1-s}+1+2^{1+\frac{1-s}{2}}\zeta\left(\frac{s-1}{2},\frac32\right).
}
The first two sums  resulted in poly-logarithms and the third in a Hurwitz zeta function.

Because of the tachyonic mode, one must choose the integration path with some care. The integral owes its convergence to the Gamma functions. These decrease in the complex $s$-plane in the directions of large imaginary and negative $s$. In the imaginary directions, from the factor $(-1)^{\frac{1-s}{2}}$ comes increase which overturns the decrease from the Gamma functions. For this reason, one should take the path as follows. First, from $-\infty$ parallel to and below the real axis, then encircling the rightmost pole in $s=3$, and continuing, to $-\infty$ above the real axis. For large negative $s$ we have a decrease from the Gamma functions and the factor $(-1)^{\frac{1-s}{2}}$ is oscillating, but not growing.
With such a choice of  path, this representation may well be used for numerical purposes.

The pole structure is determined by the Gamma functions and by the poles of  $\Sigma_N(s)$ and $\Sigma_b(s)$. For any non-integer $a$, the first expression does not have any poles. For integer $a$ it has a pole in $s=0$. The function $	\Sigma_b(s)$ has a pole in $s=3$ (independently from whether the tachyonic modes is included or not).

Representation \Ref{2.9} is very convenient for calculating the asymptotic expansion at high temperatures or small $b$. Moving the contour left across the poles, one picks up the contributions from the residua giving powers of $T/\sqrt{b}$ and the remaining integral has only lower powers in $T/\sqrt{b}$. This way we get the following expansions,
%
\eq{2.11}{\Delta_T B_4(a,b) &=\frac{2\pi^2}{3}T^4B_4(a)+i\frac{b^2}{16\pi}
\\\nn~~~&	+\frac{11b^2}{48\pi^2} \bigg[
	\ln\frac{T}{\sqrt{b}}-\gamma+\frac{6}{11}(1-\ln A+2\ln 2)
	-	\left( {\rm Li}'_0\left(e^{2\pi i a}\right)
	+{\rm Li}'_0\left(e^{-2\pi i a}\right)
	\right)\bigg]
\\\nn&~~~~~~~~~~~~~~~~~~~~~~~	+O\left(\frac{b^4}{T^4}\right),
\\\nn
\Delta_T B_2(a,b) &=\frac{T^2}{2}   B_2(a) -i\frac{b}{16\pi}
+\frac{b}{16\pi^2} \ln2
	-\frac{11b^2}{192\pi^2T^2} \left(
{\rm Li}'_{-2}\left(e^{2\pi i a}\right)+{\rm Li}'_{-2}\left(e^{-2\pi i a}\right)
\right)
	+O\left(\frac{b^2}{T^2}\right),
}
where the prime denotes differentiation with respect to the index. The structure of the expansion of $\Delta_T B_4(a,b)$, especially the missing of a contribution proportional to $T^2$ as well as further contributions odd in $T$, is due to compensations between the different gluon modes and was observed also earlier in the literature.
It must be mentioned that this expansion is not uniform in $a$ since for integer $a$ there is an additional pole in $s$ in \Ref{2.9} and some coefficients of the above expansions become singular. Thus the expansion will be different.
Accordingly, for $a=0$ the expansion takes the form,
\eq{2.12}{\Delta_T B_4(0,b) &=-\frac{\pi^2}{45}T^4
	-\frac{a_1+i}{4\pi} b^{3/2}T
	\\\nn&~~~~~~~~~-i\frac{b^2}{16\pi}	+\frac{11b^2}{48\pi^2} \bigg[
	 	\ln\frac{2\pi T}{\sqrt{b}}-\gamma+\frac{6}{11}(1-\ln A+2\ln 2)
\bigg]+O\left(\frac{b^4}{T^4}\right),
	\\\nn
\Delta_T B_2(0,b) &=\frac{T^2}{12}
	-\frac{a_2-i}{8\pi} \sqrt{b}\,T
 	+\frac{b}{16\pi^2} \ln2-i\frac{b}{16\pi}+O\left(\frac{b^2}{T^2}\right),
}
where the coefficients with the linear in $T$ terms are
\eq{2.12a}{a_1&=1-2(1-\sqrt{2})\zeta\left(-\frac12\right)
	=1-\frac{\sqrt{2}-1}{2\pi}\zeta\left(\frac32\right)
	\simeq0.828, ~~~&&\frac{a_1}{2\pi}\simeq 0.132,
\\\nn   a_2&=1-(2-\sqrt{2})\zeta\left(\frac12\right) \simeq 1.856,~~&&\frac{a_2}{2\pi}\simeq  0.148.
}
The coefficient $a_1$ was found earlier, for example $a_1
=-1+\frac{C_1}{\pi}$ in \cite{nino81-190-316}, eq. (3.7), $a_1=\frac{2\pi}{3}a$ in \cite{eber98-13-1723},  eq. (30), and $a_1=C_2$ in \cite{meis02-66-105006}, eq. (C4).
Each of the expansions in \Ref{2.12} includes  terms linear in $T$.
Their real parts originate from the non-tachyonic modes and their imaginary parts from the tachyonic mode.

Finally we add  the zero temperature part.   It contributes to the order $b^2$. For $a\ne 0$, adding \Ref{2.1.6} to \Ref{2.11}, we arrive at
\eq{2.13}{ B_4(a,b) &=\frac{2\pi^2}{3}T^4B_4(a)
		+\frac{11b^2}{48\pi^2} \bigg[ \ln(2T)-\gamma
	-	{\rm Li}'_0\left(e^{2\pi i a}\right)-{\rm Li}'_0\left(e^{-2\pi i a}\right)
	\bigg]
	+O\left(\frac{b^4}{T^4}\right),
	\\\nn
B_2(a,b) &=\frac{T^2}{2} B_2(a)
	-\frac{11b^2}{192\pi^2T^2} \left(
	{\rm Li}'_{-2}\left(e^{2\pi i a}\right)+{\rm Li}'_{-2}\left(e^{-2\pi i a}\right)
	\right)  +O\left(\frac{b^2}{T^2}\right),
}
Here, the imaginary parts  and $\ln b$ canceled (see the remark after eq. \Ref{2.17}).  For $a=0$, adding \Ref{2.1.6} to \Ref{2.12},  we get
\eq{2.13a}{ B_4(0,b) &=-\frac{\pi^2}{45}T^4
	-\frac{a_1+i}{4\pi} b^{3/2}T
	+\frac{11b^2}{48\pi^2} \bigg[
	\ln(4\pi T)-\gamma
	\bigg]+O\left(\frac{b^4}{T^4}\right),
	\\\nn
 B_2(0,b) &=\frac{T^2}{12}
	-\frac{a_2-i}{8\pi}\sqrt{b}\, T
 +O\left(\frac{b^2}{T^2}\right).
}
Again, the $\ln b$-term was canceled; but an imaginary part remained. We mention that the tachyonic mode is included in the above formulas. It is to be mentioned that there is no imaginary part in \Ref{2.13}. Sometimes in literature, this was attributed to the finite temperature. However, in general, at finite temperature, and especially in the limiting case \Ref{2.13a}, there is an imaginary part.

The main advantage of the current approach is in the combination of the double sums in eq.  \Ref{2.8} with the Mellin representation, \Ref{Me}, which allows for a machined evaluation of the residua.

We add the following remark. The above expansions are for $a$ and $b$ fixed, $T\to\infty$. One might wish to consider $A_0$ fixed in place of $a$. With \Ref{ab}, this implies $a\to 0$. The variable $a$ appears only in the argument of the $\cos$ in \Ref{2.10}. An expansion would result in growing powers, $N^{2k}$, ($k=0.1.2.\dots$), of $N$. This would produce factors $\zeta(s-2k+1)$ and add  poles at $s=2k$ in \Ref{2.9}. One would be forced to move the integration contour right, where the integrand is growing because of the Gamma functions. Convergence would be lost and the results, which are of order $(gA_0/\sqrt{b})^{2k}$ (all contribute with the same power of $T$), would not be reliable. Moreover, summing the resulting series (this can be done explicitly), would come in conflict with the expansion \Ref{2.13a}.


\subsection{\label{T2.3}Representation of the non-tachyonic part in terms of Theta functions}
In order to get  formulas that would be useful for numerical  calculation of $B_4$ and $B_2$ in parameter regions where the sum representations \Ref{2.8} are not effective, we consider a representation in terms of theta functions. Such a representation was used,  for example, in \cite{star94-322-403}. For this, we must do the split \Ref{2.4} into non-tachyonic and tachyonic parts and in this subsection, we consider the first one.

We return to the representation \Ref{2.2} and use the  Jacobi theta function,
\eq{7}{ \Theta_3(z,q)=\sum_{N=-\infty}^{\infty}q^{N^2}\cos(2 z N),
	~~~~~\frac12(\Theta_3(z,q)-1) = \sum_{N=1}^\infty q^{N^2} \cos(2 z N),
}
with $q=\exp\left(-\frac{1}{4T^2t}\right)$ and $z=\pi a$, and using also the last line in \Ref{3} and \Ref{2.1.0}, we arrive at the representations
\eq{8}{\Delta_T  B_4^{nt}(a,b) &=-\frac{b}{16\pi^2}\int_0^\infty\frac{dt}{t^2}
	\left[\Theta_3\left(\pi a,
	e^{-\frac{1}{4T^2t}}\right)-1\right] \ e^{-tb} \coth(tb),
\\\nn
\Delta_T  B_3^{nt}(a,b) &=-\frac{bT^{-1}}{64\pi^2}\int_0^\infty\frac{dt}{t^2}
\ \Theta'_3\left(\pi a,
e^{-\frac{1}{4T^2t}}\right) \ e^{-tb} \coth(tb),
\\\nn	\Delta_T  B_2^{nt}(a,b) &=\frac{b}{16\pi^2}\int_0^\infty\frac{dt}{t}
	\left[\Theta_3\left(\pi a,
	e^{-\frac{1}{4T^2t}}\right)-1\right] \ e^{-tb} \coth(tb).
\\\nn	\Delta_T  B_1^{nt}(a,b) &=-\frac{bT^{-1}}{64\pi^2}\int_0^\infty\frac{dt}{t}
\ \Theta'_3\left(\pi a,
e^{-\frac{1}{4T^2t}}\right) \ e^{-tb} \coth(tb).
}
The subtraction of the '$1$' in the square bracket is because of the dropped  $N=0$-term. In the odd number functions, this term is absent.
In these formulas,  we also carried out the sum over the Landau levels and over the spin with the exception, as announced,  for the tachyonic mode, $n=0,\ \sig=-2$.

\subsection{\label{T2.4}The tachyonic part}
For the tachyonic part, we consider three representations, one is a sum over the Matsubara frequencies, and the other uses the Abel-Plana formula and the third is in terms of Hankel functions.

\subsubsection{\label{T2.4.1}Sum representation}
We start with the sum  representation. We go back to representation \Ref{2.1} where   only the integration over $k_3$ is carried out, and include only the tachyonic mode,
\eq{2.14}{ B_4^{ta}(a,b) &=-\pa_s\int_0^\infty\frac{dt}{t}\,\frac{t^{s-\frac12}}{\sqrt{4\pi}\Gamma(s)}
	T\sum_\ell \frac{b}{4\pi}
	\exp\left\{-t\left[\left(2\pi T(\ell+a)\right)^2-b-i0\right]\right\},
	\\\nn
	B_2^{ta}(a,b) &=\int_0^\infty\frac{dt}{t}\,\frac{t^{s+\frac12}}{\sqrt{4\pi}\Gamma(s+1)}
	T\sum_\ell \frac{b}{4\pi}
	\exp\left\{-t\left[\left(2\pi T(\ell+a)\right)^2-b-i0\right]\right\}.
}
Next we carry out the integration over $t$,
\eq{2.15}{ B_4^{ta}(a,b) &=-\pa_s \frac{ \Gamma(s-\frac12)}{\sqrt{4\pi}\Gamma(s)}
	\frac{bT^2}{2(2\pi T)^{2s} }
\	\sum_\ell \left((\ell+a)^2-\beta\right)^{\frac12-s},
	\\\nn
	B_2^{ta}(a,b) &= \frac{\Gamma(s+\frac12)}{\sqrt{4\pi}\Gamma(s+1)}
 	\frac{b}{8\pi^2(2\pi T)^{2s} }
\  \sum_\ell \left((\ell+a)^2-\beta\right)^{-\frac12-s},
}
where we defined  $\beta=\frac{b}{(2\pi T)^2}+i0$. To proceed, we must perform the analytic continuation to $s=0$.

We start with the imaginary part. Since only a finite number of terms enters  one can put $s=0$ and arrives at
\eq{2.16}{ \Im B_4^{ta}(a,b) &=
-	\frac{b T}{4\pi} \left(\sum_{\ell=0}^{[\sqrt{\beta}-a]}\sqrt{b-(2\pi T (\ell+a))^2}
		+	\sum_{\ell=1}^{[\sqrt{\beta}+a]}\sqrt{b-(2\pi T (\ell-a))^2}\right),
\\\nn  \Im B_2^{ta}(a,b) &=
	\frac{bT }{8\pi} \left(\sum_{\ell=0}^{[\sqrt{\beta}-a]}
	\frac{1}{\sqrt{b-(2\pi T (\ell+a))^2}}
+	\sum_{\ell=1}^{[\sqrt{\beta}+a]}
	\frac{1}{\sqrt{b-(2\pi T (\ell-a))^2}}\right).
}
There is an imaginary part if either $\sqrt{b}>\pi T$, or, if $\sqrt{b}<\pi T$, the relations  $2\pi Ta<\sqrt{b}$	or  $2\pi T(1-a)<\sqrt{b}$ hold. For $a=0$ there is always an imaginary part.
There is no imaginary part if
\eq{2.17}{ \sqrt{b}< 2\pi T a%
}
holds. For instance, for high $T$, with $a$ fixed, there is no imaginary part as can be seen from the corresponding asymptotic expansions in \Ref{2.13}, whereas in the expansion \Ref{2.13a} for $a=0$ there is an imaginary part. 

For the real part we proceed as follows.
First, we separate the contribution from $\ell=0$. Then we  add and subtract the first  terms of the expansion for large $\ell$. This way we represent,
\eq{2.18a}{ B_4^{ta}(a,b) &= \frac{bT^2}{2}V_4,
	&B_2^{ta}(a,b) &= \frac{b}{16\pi^2}V_2,
}
with  three contributions,
\eq{2.18}{V_\nu=V_{\nu,0}+V_{\nu,1}+V_{\nu,2},~~~(\nu=4,2).
}
The first ones are
\eq{2.19}{V_{4,0}&= \sqrt{a^2-\beta},
	~~~&V_{2,0}&= \frac{1}{\sqrt{a^2-\beta}}.
}
In the  second ones, the sum over $\ell$ results in Riemann zeta functions,
\eq{2.20}{ V_{4,1} &= 2\zeta(2s-1)+(2s-1)(2s a^2+\beta)\zeta(2s+1),
	\\\nn  V_{2,1} &=2\zeta(2s+1)+(2s+1)(2(s+1)a^2+\beta)\zeta(2s+3).
}
The last terms are converging sums and we could put $s=0$,
\eq{2.21}{V_{4,2} &= \sum_{\ell=1}^{\infty}
	\left(\sqrt{(\ell+a)^2-\beta}+\sqrt{(\ell-a)^2-\beta}
	-\left(2\ell-\frac{\beta}{\ell}\right) \right),
\\\nn V_{2,2} &= \sum_{\ell=1}^{\infty}
\left(\frac{1}{\sqrt{(\ell+a)^2-\beta}}+\frac{1}{\sqrt{(\ell-a)^2-\beta}}
-\frac{2}{\ell}  \right).
}
Finally, inserting into \Ref{2.18} and performing the limit $s=0$ we arrive at
\eq{2.22}{ B_4^{ta}(a,b) &=\frac{bT^2}{2} V_{4,0}
	+\frac{bT^2}{2} \left(-\frac16 - a^2 \right)+ \frac {b^2}{8\pi^2}(\ln(4\pi T)-\gamma)+\frac{bT^2}{2}V_{4,2},
\\\nn	B_3^{ta}(a,b) &=  \frac{bT}{8\pi}\sum_{\ell=0}^\infty
		\left(\frac{\ell+a}{\sqrt{(\ell+a)^2-\beta}}
				-\frac{\ell+1-a}{\sqrt{(\ell+1-a)^2-\beta}}   \right),
\\\nn	 B_2^{ta}(a,b) & = \frac{b}{16\pi^2}(V_{2,0}+V_{2,2})
-\frac{b}{8\pi^2}(\ln(4\pi T) -\gamma)
,
\\\nn	 B_1^{ta}(a,b) & = \frac{b}{64\pi^3T}  \sum_{\ell=0}^\infty
			\left( \frac{\ell+a}{((\ell+a)^2-\beta)^{3/2}}
					- \frac{\ell+1-a}{((\ell+1-a)^2-\beta)^{3/2}}  \right).
}
In these formulas, $\gamma$ is Euler's constant.
We mention that we dropped a divergent piece from $ B_2^{ta}(a,b) $ which does not depend on $T$. For $ B_3^{ta}(a,b) $ and $ B_1^{ta}(a,b) $ we used \Ref{3}. In these, the sums are convergent.

We mention that in this representations the imaginary parts \Ref{2.16} are included. In case, one is interested only in the real parts, one may simply calculate \Ref{2.22} and take the real part.

\subsubsection{\label{T2.4.2}Integral representation}
In this subsection, we derive an integral representation for the contribution from the tachyonic mode which is alternative to the sum representation in the preceding subsection. For this, we go back to eqs. \Ref{2.1}, keep only the tachyonic mode and carry out the integration over $t$,
\eq{2.23}{ B_4^{ta}(a,b) &=-\pa_s   \frac{\Gamma(s-\frac12)}{\sqrt{4\pi}\Gamma(s)}
	 \frac{bT}{4\pi}\sum_\ell
\left(\left(2\pi T(\ell+a)\right)^2-b-i0\right)^{\frac12-s},
	\\\nn
	B_2^{ta}(a,b) &= \frac{\Gamma(s+\frac12)}{\sqrt{4\pi}\Gamma(s+1)}
	\frac{bT}{4\pi}\sum_\ell
	\left(\left(2\pi T(\ell+a)\right)^2-b-i0\right)^{-\frac12-s}.
}
The sum over $\ell$ can be transformed into  integrals using the Abel-Plana formula which, generalized to the inclusion of $a$, takes the form
\eq{AP}{T\sum_\ell \left(\left(2\pi T(\ell+a)\right)^2+b_0\right)^{\frac12-s}
	=\int\frac{dk}{2\pi}\left(k^2+b_0\right)^{\frac12-s}
	-\frac{\cos(\pi s)}{\pi}\int_{\sqrt{b_0}}^\infty d\om\,\left(\om^2-b_0\right)^{\frac12-s}
	h\left(a,\frac{\om}{T}\right),
}
where we defined
\eq{2.24}{h(a,\om)&=\frac{1}{e^{2\pi i a+\om}-1}+\frac{1}{e^{-2\pi i a+\om}-1}
	=\frac{ \cos(2\pi a)-e^{-\om}}{\cosh(\om)-\cos(2\pi a)}.
}
Also we will need the function
\eq{2.25}{g(a, \om)&=\frac{i}{e^{2\pi i a+\om}-1}+\frac{-i}{e^{-2\pi i a+\om}-1}
	=\frac{\sin(2\pi a)}{\cosh(\om)-\cos(2\pi a)},
}
and we mention the relation $\frac{1}{2\pi}\pa_ah(a,\om)=\pa_\om g(a,\om)$.
Here we introduced the parameter $b_0$ which is assumed to be real. To apply this formula to \Ref{2.23}, we need to do the analytic continuation
\eq{2.26}{b_0=-b-i0~~~\mbox{and}~~~\sqrt{b_0}=-i\sqrt{b}.
}
On the right side of eq. \Ref{AP}, the first integral gives the zero temperature part and the second the temperature-dependent part which will be marked by '$\Delta_T$' as before.
The zero-temperature parts are
\eq{2.27}{ B_4^{ta,T=0}(a,b) &=-\pa_s   \frac{\Gamma(s-\frac12)}{\sqrt{4\pi}\Gamma(s)}
	\frac{b}{4\pi}\int\frac{dk}{2\pi}\left(k^2+b_0\right)^{\frac12-s}
=-i\frac{b^2}{16\pi}+\frac{b^2}{16\pi^2}\left(-1+\ln b\right),
\\\nn
 B_2^{ta,T=0}(a,b) &=  \frac{\Gamma(s+\frac12)}{\sqrt{4\pi}\Gamma(s+1)}
\frac{b}{4\pi}\int\frac{dk}{2\pi}\left(k^2+b_0\right)^{-\frac12-s}
=\frac{b}{16\pi^2 s}+i\frac{b}{16\pi}-\frac{b}{16\pi^2}\ln b.
}
These expressions constitute the tachyonic part of \Ref{2.16}.

For the temperature-dependent parts, we arrive at the expressions
\eq{2.28}{ \Delta_T B_4^{ta}(a,b) &= -\frac{bT^2}{4\pi^2}\int_{\sqrt{\beta_0}}^\infty
	d\om\,\sqrt{\om^2-\beta_0}\ h\left(a,\om\right),
\\\nn   \Delta_T B_3^{ta}(a,b) &= \frac{bT}{8\pi^2}\int_{\sqrt{\beta_0}}^\infty
d\om\,\frac{\om}{\sqrt{\om^2-\beta_0}}\ g\left(a,\om\right).
\\\nn   \Delta_T B_2^{ta}(a,b) &= \frac{b}{8\pi^2}\int_{\sqrt{\beta_0}}^\infty
d\om\,\frac{1}{\sqrt{\om^2-\beta_0}}\ h\left(a,\om\right).
\\\nn   \Delta_T B_1^{ta}(a,b) &= -\frac{bT^{-1}}{16\pi^2}\int_{\sqrt{\beta_0}}^\infty
d\om\,\frac{1}{\sqrt{\om^2-\beta_0}}\ \pa_\om g\left(a,\om\right).
}
Since here there  are no divergences we could put $s=0$. Also, we made the substitution $ \om\to\om T$ and introduced the notation $\beta_0=\frac{b_0}{T^2}$ for notation convenience. In the second line, we integrated by parts. In the last line, we did not integrate by parts to avoid a singularity at the lower integration boundary.

Now we do the analytic continuation \Ref{2.26}, which here takes the form $\sqrt{\beta_0}=-i\sqrt{\beta}$, $\beta=b/T^2$.
Thereby the beginning of the integration path moves in the complex $\om$-plane down to the negative imaginary axis. Numerical integration in such formulas is possible (the integration does converge). For practical purposes one may substitute $\om=-i\sqrt{\beta}+s$ with integration over real $s=0,\dots,\infty$ and taking the real part of the integral (the imaginary parts are given by \Ref{2.16}).

There is an alternative to the integration in the complex plane.
We may deform the integration path to run from $-i\sqrt{\beta}$ to the origin along the imaginary axis, and from the origin to infinity along the real axis. Accordingly, we split
\eq{2.29}{\Delta_T  B_4^{ta}(a,b) &=T_1+T_2,~~~~~\Delta_T  B_2^{ta}(a,b)=U_1+U_2.
}
Before carrying out this program it is useful to integrate by parts.
Representing the right side of \Ref{2.24} in the form
\eq{2.30}{ h(a,\om) &= \pa_\om f(a,\om)
}
with
\eq{2.31}{ f(a,\om) &= \ln\left(1+e^{-2\om}-2e^{-\om}\cos(2\pi a)\right)
	=-\om+\ln\left[2\left(\cosh(\om)-\cos(2\pi a)\right)\right],
}
we arrive at
\eq{2.32}{  \Delta_T B_4^{ta}(a,b) &= \frac{bT^2}{4\pi^2}\int_{\sqrt{\beta_0}}^\infty
	d\om\,\frac{\om}{\sqrt{\om^2-\beta_0}}\ f\left(a,\om\right),
	\\\nn   \Delta_T B_2^{ta}(a,b) &= \frac{b}{4\pi^2}\int_{\sqrt{b_0}}^\infty
	d\om\,\frac{\om}{\left(\om^2-\beta_0\right)^{3/2}}\
	\left(f\left(a,\om\right)-f\left(a,\sqrt{\beta_0}\right)\right).
}
There are no surface terms.

In the path along the imaginary axis we substitute $\om=-i x$ and get for the real parts of the  integrals
\eq{2.33}{ T_1 &= \frac{bT^2}{4\pi^2}\int_0^{\sqrt{\beta}}\frac{dx\,x}{\sqrt{\beta-x^2}}
	\ln\left(2 \left|\cos(x)-\cos(2\pi a)\right|\right),
	\\\nn
	U_1 &= \frac{b}{4\pi^2}
	\int_0^{\sqrt{\beta}}\frac{dx\,x}{(\beta-x^2)^{3/2}}
	\ln\left|\frac{  \cos(x)-\cos(2\pi a) }{ \cos(\sqrt{\beta})-\cos(2\pi a)}\right|.
}
The absolute values in the logarithms let us get the contributions from the real parts (the imaginary parts were considered above in eq. \Ref{2.16}). Finally, we make the substitution $x\to x\sqrt{\beta}$ and arrive at
\eq{2.34}{ T_1 &= \frac{b^{3/2}T}{4\pi^2}\int_0^{1}\frac{dx\,x}{\sqrt{1-x^2}}
	\ln\left(2 \left|\cos\left(x\frac{\sqrt{b}}{T}\right)-\cos(2\pi a)\right|\right),
	\\\nn
	U_1 &= \frac{b^{1/2}T}{4\pi^2}
	\int_0^{1}\frac{dx\,x}{(1-x^2)^{3/2}}
	\ln\left|\frac{  \cos\left(x\frac{\sqrt{b}}{T}\right)-\cos(2\pi a) }{ \cos\left(\sqrt{\frac{\sqrt{b}}{T}}\right)-\cos(2\pi a)}\right|.
}
For the contributions from the path along the real axis we simply use \Ref{2.32} to get
\eq{2.35}{T_2 &= \frac{bT^2}{4\pi^2}\int_0^\infty\frac{d\om\,\om}{\sqrt{\om^2+\beta}}
	f(a,\om),
	\\\nn
	U_2 &= -\frac{b}{4\pi^2}\int_0^\infty \frac{d\om\,\om}{(\om^2+\beta)^{3/2}}
	\left(f(a,\om)-f\left(a,-i\sqrt{\beta}\right)\right),
	\\\nn &=
	-\frac{b}{4\pi^2}\ln\left(2\left|\cos(\beta)-\cos(2\pi a)\right|\right)+
	\frac{b}{4\pi^2}\int_0^\infty \frac{d\om\,\om}{(\om^2+\beta)^{3/2}}
	f(a,\om) .
}
In the last line another imaginary piece was dropped.
Doing the substitution $x\to x\sqrt{\beta}$ we arrive finally at
\eq{2.36}{T_2 &= \frac{b^{3/2}T}{4\pi^2}\int_0^\infty\frac{d\om\,\om}{\sqrt{\om^2+1}}\
	f\left(a,\om\frac{\sqrt{b}}{T}\right),
	\\\nn
	U_2 &=
	-\frac{b}{4\pi^2}\ln\left(2\left|\cos\left(\frac{\sqrt{b}}{T}\right)-\cos(2\pi a)\right|\right)+
	\frac{b^{1/2}T}{4\pi^2}\int_0^\infty \frac{d\om\,\om}{(\om^2+1)^{3/2}}\
	f\left(a,\om\frac{\sqrt{b}}{T}\right) .
}

\subsubsection{\label{T2.4.3}Representation of the tachyonic part in terms of Bessel functions and the low-T expansion}
Although in Sect. \ref{T2.2} it was observed that a representation of the tachyonic mode in terms of Bessel functions is less convenient for numerical evaluation, it is convenient at small temperature (or large magnetic field) as mentioned, for example, in \cite{star94-322-403}. For this reason we provide it here. We mention that the low-temperature expansion of the non-tachyonic part is exponentially small, as can be seen from \Ref{2.8} because only the modified Bessel functions enter. We start with the representation \Ref{2.8} and take only the tachyonic mode, $n=0,~\sig=-2$, such that $\sqrt{2n+1+\sig}=-i$, where the analytic continuation was done as discussed in the beginning of section \ref{T2}. With the standard properties of the Bessel functions, we get from \Ref{2.8}
\eq{2.37}{\Delta_TB_4^{ta}&=-i\frac{b^{3/2}T}{4\pi}
	\sum_{N=1}^\infty\frac{cos(2\pi aN)}{N}
	H_1^{(1)}\left(\frac{\sqrt{b}}{T}N\right),
\\\nn
\Delta_TB_2^{ta}&=i\frac{b}{8\pi}	
\sum_{N=1}^\infty{cos(2\pi aN)}
H_0^{(1)}\left(\frac{\sqrt{b}}{T}N\right),
}
where $H_\nu^{(1)}(z)$ are Hankel functions. Using their leading order asymptotic expansion for large arguments,
\eq{2.38}{H_\nu^{(1)}(z) &\simeq \sqrt{\frac{2}{\pi z}}\ e^{i\om} ~~~\mbox{with}~~~\om=z-\frac{\pi}{2}\nu-\frac{\pi}{4},
}
we arrive at
\eq{2.39}{ \Delta_TB_4^{ta}&\underset{T\to 0}{\simeq}
		-\frac{b^{5/4}T^{1/2}}{2 (2\pi)^{3/2}}
		\left(
		{\rm Li}_\frac32\left(e^{i2\pi a+i \frac{\sqrt{b}}{T}}\right)+
		{\rm Li}_\frac32\left(e^{-i2\pi a+i \frac{\sqrt{b}}{T}}\right)\right)\, e^{-i\frac{\pi}{4}},
\\\nn
 \Delta_TB_2^{ta}&\underset{T\to 0}{\simeq}
\frac{b^{3/4}T^{3/2}}{2 (2\pi)^{3/2}}
\left(
{\rm Li}_\frac12\left(e^{i2\pi a+i \frac{\sqrt{b}}{T}}\right)+
{\rm Li}_\frac12\left(e^{-i2\pi a+i \frac{\sqrt{b}}{T}}\right)\right)\, e^{i\frac{\pi}{4}}.
}
As can be seen, this first low-temperature correction to \Ref{2.1.6}  has, as a function of $a$, a structure similar  the high-temperature expansion \Ref{2.11}  with the difference of a fast rotation (in the complex plane) of the arguments of the poly-logarithms.

\subsection{\label{T2.5}The complete expressions}
The complete expressions for the real parts of the functions $B_x(a,b)$, defined in \Ref{3}, follow from  \Ref{2.3} and \Ref{2.4}. For the $T=0$-part we have \Ref{2.1.6} and for the  non-tachyonic part either \Ref{2.8} or \Ref{8}. For the tachyonic part we have either \Ref{2.22} or \Ref{2.28} with \Ref{2.29}.
Together the real parts read
\eq{c1}{\Re B_n(a,b) &=
\Re \left(	B_n^{T=0}(a,b)+ 		\Delta_T 	B_n^{nt}(a,b)+	\Delta_T 	B_n^{ta}(a,b)
\right),
}
for $n=1,\dots,4$.
The imaginary parts are given by eq. \Ref{2.16}.

For $SU(2)$, the complete one and two-loop expression for the effective action is given by \Ref{W2}.
\eq{W2a}{ W^{SU(2)}_{gl} &=\frac{b^2}{2g^2}+B_4(0,0)+2B_4\left(a,b\right)
	\\\nn&~~~	+\frac{g^2}{2}\left[
	B_2\left(a,b\right)^2
	+2 B_2\left(0,b\right) B_2\left(a,b\right)
-8(1-\xi) B_3\left(a,b\right)B_1\left(a,b\right)
	\right]
}
where we added the tree contribution.

\section{\label{T3}The minimum of the effective action in the pure magnetic case ($A_0=0$)}
In this section, we apply the results of the preceding section to calculate the effective action as a function of the magnetic background field  $H$ and to find its minimum.
\subsection{\label{T3.1}The case $T=0$}
We start from the zero temperature case for completeness, although in QCD this case is not physical in the perturbative approach. Also, it will serve to check  the numerical investigations at finite, but low temperatures. In the zero temperature case we have $2\pi T(\ell+a)\to k_4$ in \Ref{3} and the functions $B^{T=0}_x(a,b)$ in \Ref{2.3} do not have any dependence on the $A_0$-background due to the translational invariance in $k_4$. These functions were calculated in subsection \ref{T2.1} and we use eq. \Ref{2.1.5}.

Insertion of \Ref{2.1.6} into  \Ref{W2a} delivers
\eq{3.2}{W_{gl}^{SU(2)}  &=\frac{b^2}{2g^2}
	+\frac{11\, b^2}{48\pi^2}\left(\ln\frac{b}{\mu^2}-\frac12\right)-i\frac{b^2}{8\pi^2}
	+g^2\frac{\ln^2(2)\ b^2}{128\pi^4}.
}
Here, the first term is the classical energy of the background. The second term is the famous vacuum energy of $SU(2)$ in magnetic background \cite{savv77-71-133}. In QED, the first and the second term (with different coefficients) form the Euler-Heisenberg Lagrangian. The third term is the known imaginary part, causing the instability on the one-loop level. The last term is the contribution from the second loop, which was  in this context never considered since it is expected to give only a small contribution.

The energy \Ref{3.2} has a non-trivial minimum resulting from the logarithmic term,
\eq{3.3}{ b_{min}&=\mu^2 \, e^{-\frac{24\pi^2}{11g^2} -\frac{3\ln^2(2)}{98\pi^2}g^2}
	=\mu^2 \, e^{-\frac{24\pi^2}{11g^2}}
	\left(1-\frac{3\ln^2(2)\ g^2}{88\pi^2} +\dots\right),
\\\nn	W^{SU(2)}_{min} &=-\frac{11\mu^4}{96\pi^2}\
	e^{-\frac{24\pi^2}{11g^2}   -\frac{3\ln^2(2)}{44\pi^2}g^2   }
= -\frac{11\mu^4}{96\pi^2}\ e^{-\frac{24\pi^2}{11g^2}}
		\left(1-\frac{3\ln^2(2)\ g^2}{44\pi^2}+\dots\right).
}
The first line is the field in the minimum, i.e., the condensate, and the second line is the energy in this minimum. The second loop appears as an additional term in the exponentials and its expansion in order $g^2$ is shown in the parenthesizes.
We mention the known feature that this minimum is non-perturbative. It has an essential singularity at $g=0$ from the exponential prefactor. Also, it depends essentially on the arbitrary scale $\mu$, which is the only dimensional parameter in this case.

\begin{figure}[h]
	\includegraphics[width=0.65\textwidth]{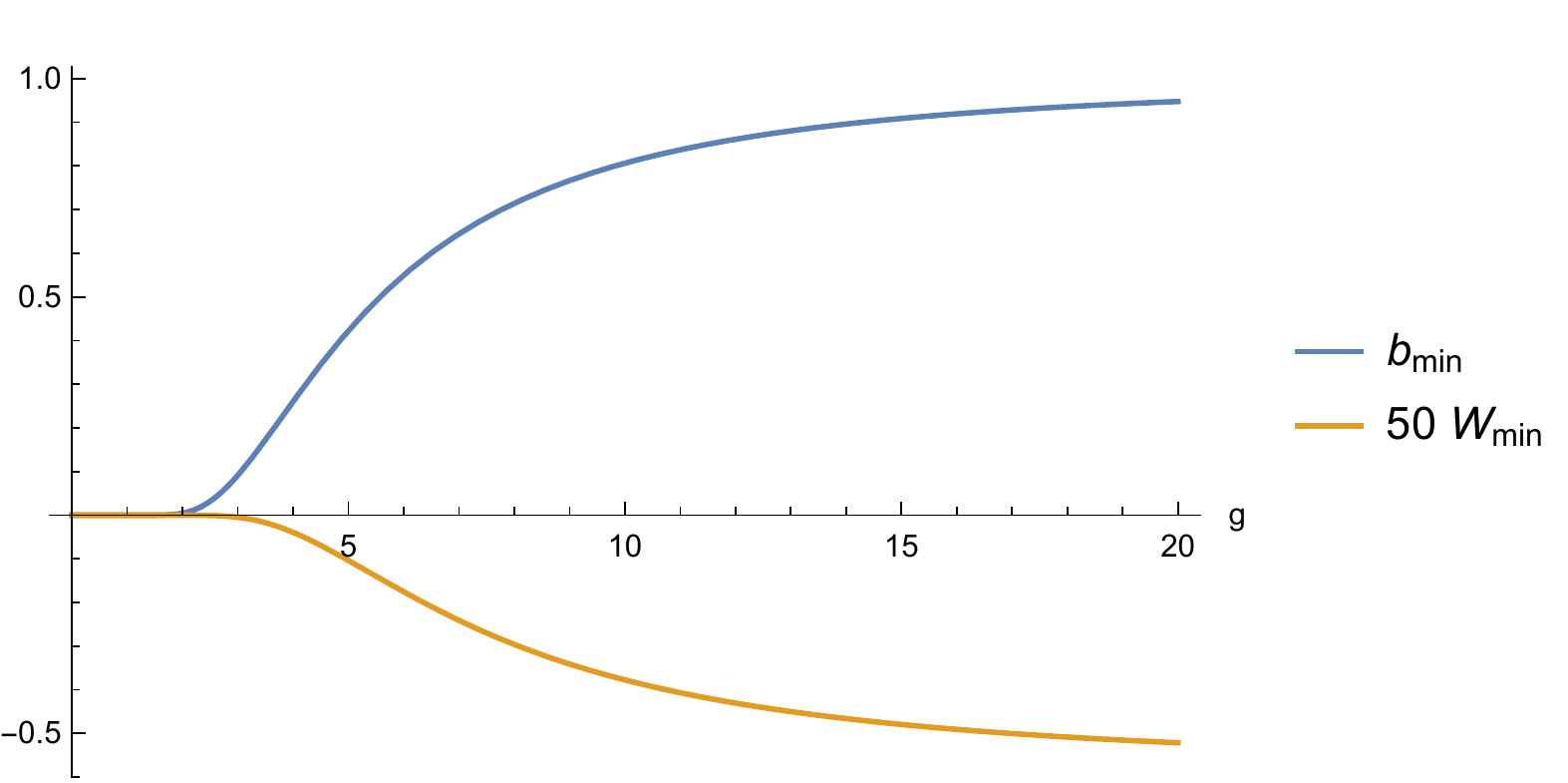}
 	\caption{The dependence of the vacuum energy in one-loop order and of the condensate field \Ref{3.3}   on the coupling constant $g$ for $SU(2)$. The vacuum energy is amplified by a factor of 50 to fit together with the field in one plot.}
	\label{fig:1}\end{figure}

In one-loop order, i.e., without the second term in the exponentials, or without the  parenthesis, the behavior of these expressions is well known. For small coupling $g$ they vanish fast, for larger coupling they go into saturation. These features are illustrated in   Fig. \ref{fig:1}.

\begin{figure}[h]
	\includegraphics[width=0.5\textwidth]{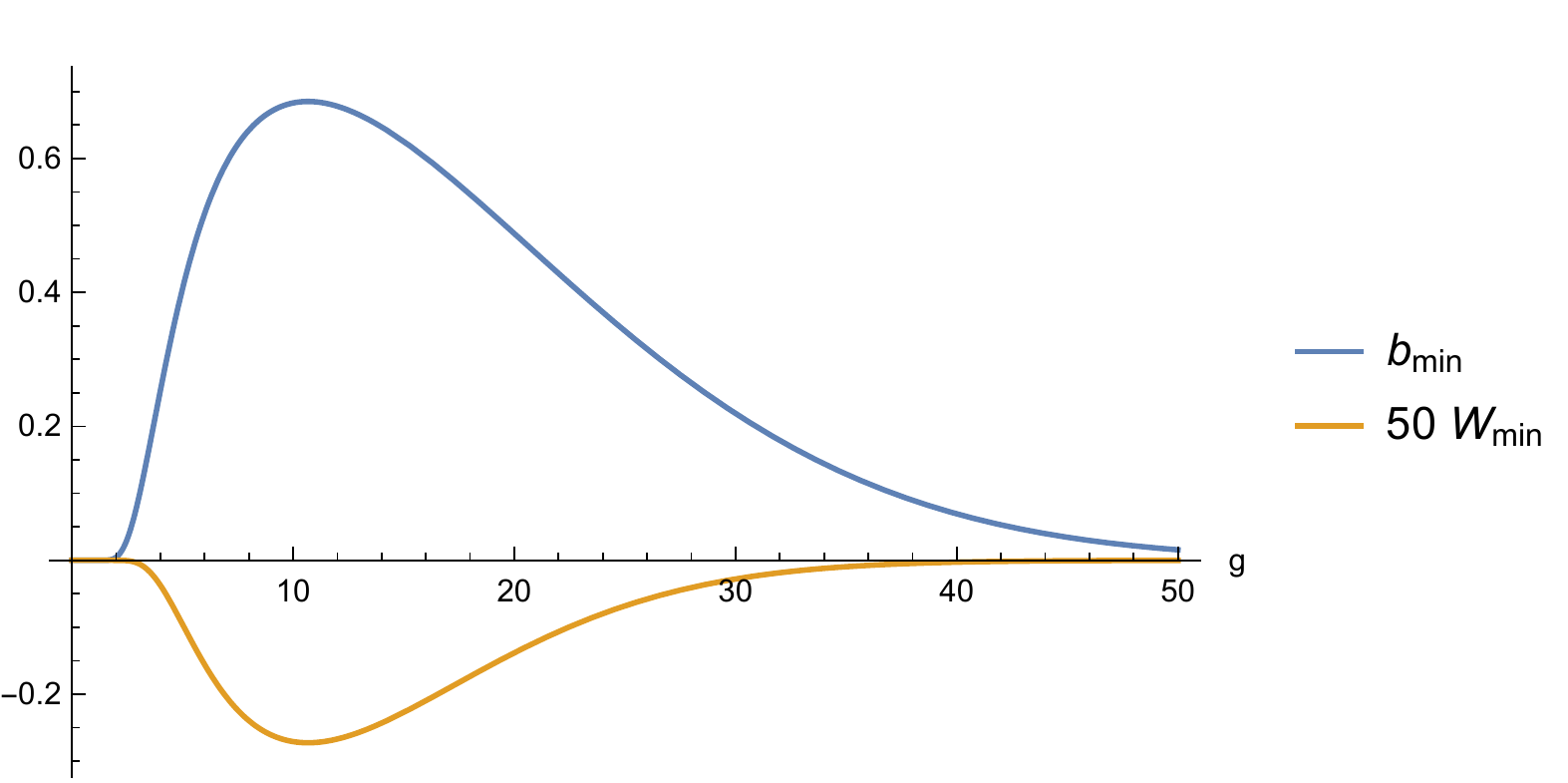}
		\includegraphics[width=0.5\textwidth]{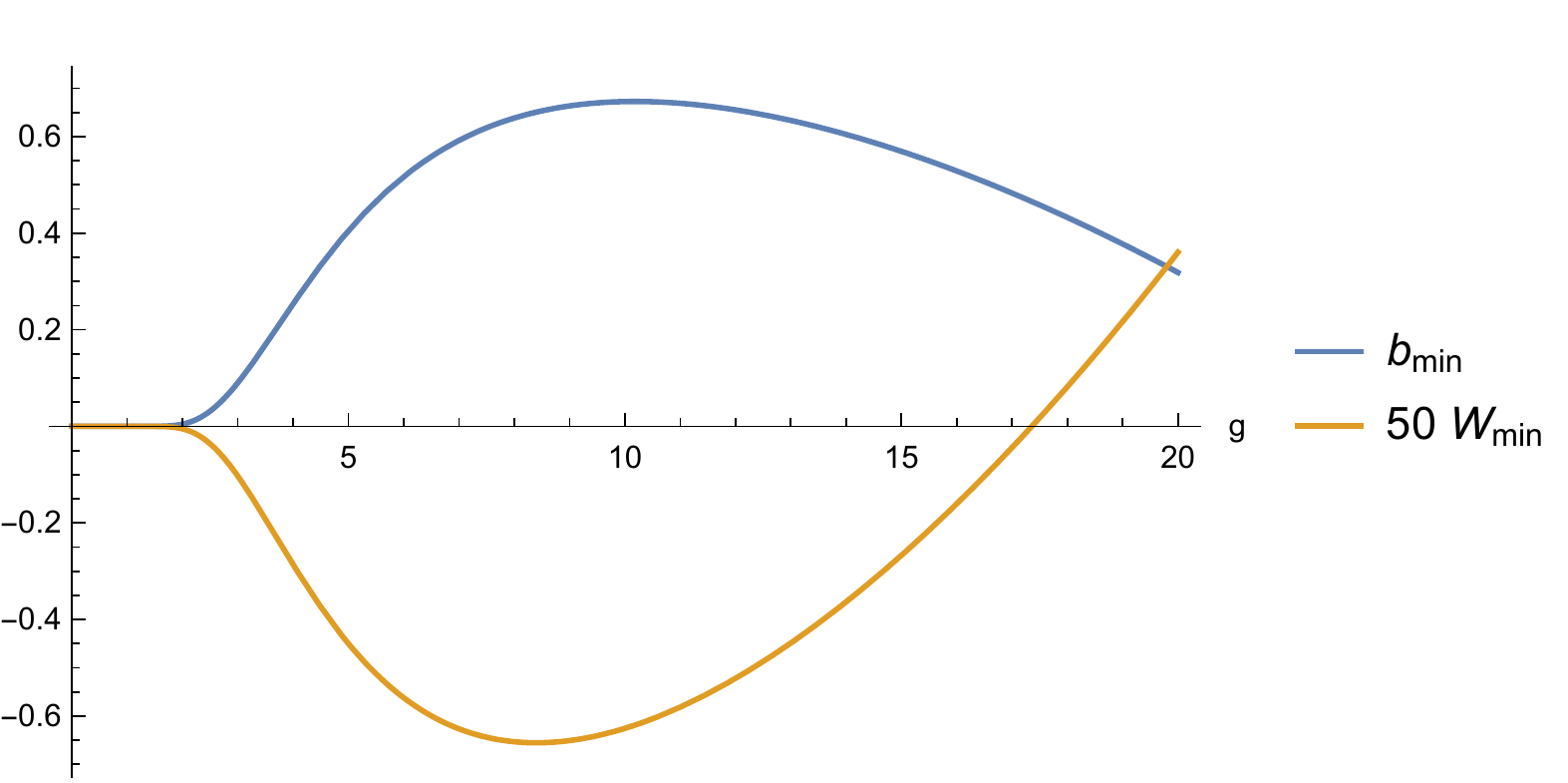}
	\caption{The dependence of the vacuum energy in two-loop order and of the condensate field \Ref{3.3}   on the coupling constant $g$ for $SU(2)$. In the left panel we show the complete expression (middle terms in \Ref{3.2}) and in the right panel we show the corresponding expansion (last terms in \Ref{3.3}).
	The vacuum energy is amplified by a factor of 50 to fit together with the field in one plot.}
	\label{fig:2}\end{figure}

In two-loop order, the picture changes for large $g$. The additional contribution compensates   the negative energy from the one-loop approximation and the minimum disappears.  We demonstrate this feature in Fig. \ref{fig:2}. This way, at larger coupling, the second loop is not  a small  addendum and demonstrates the breakdown of the perturbative expansion.

\subsection{\label{T3.2}The high-$T$ case}
In this subsection, we consider the case of high temperature. We use the formulas \Ref{2.13a} from Section \ref{T2.2} and insert these into \Ref{W2a}. With \Ref{ab}, the effective potential reads
\eq{3.6}{W^{SU(2)}_{gl}&=
	 \frac{b^2}{2 g^2}-\frac{\pi ^2 T^4}{15}
	 -\frac{a_1 b^{3/2} T}{2 \pi }
	 +\frac{11 b^2( \log (4 \pi  T)-\gamma)}{24 \pi ^2}
	 +g^2\left( \frac{T^4}{24}-\frac{{a_2} \sqrt{b} T^3}{12 \pi }
	 +\frac{{a_2}^2 b T^2}{32 \pi ^2}
	 \right)
}
Again, the first term is the classical energy. The   terms proportional to $T^4$ constitute the gluon black body radiation and there is no term logarithmic in $b$. The contribution from the second loop is in the parenthesis. It has a $T^3$-contribution which was missing in the one-loop part.

\begin{figure}[h]
	\includegraphics[width=0.5\textwidth]{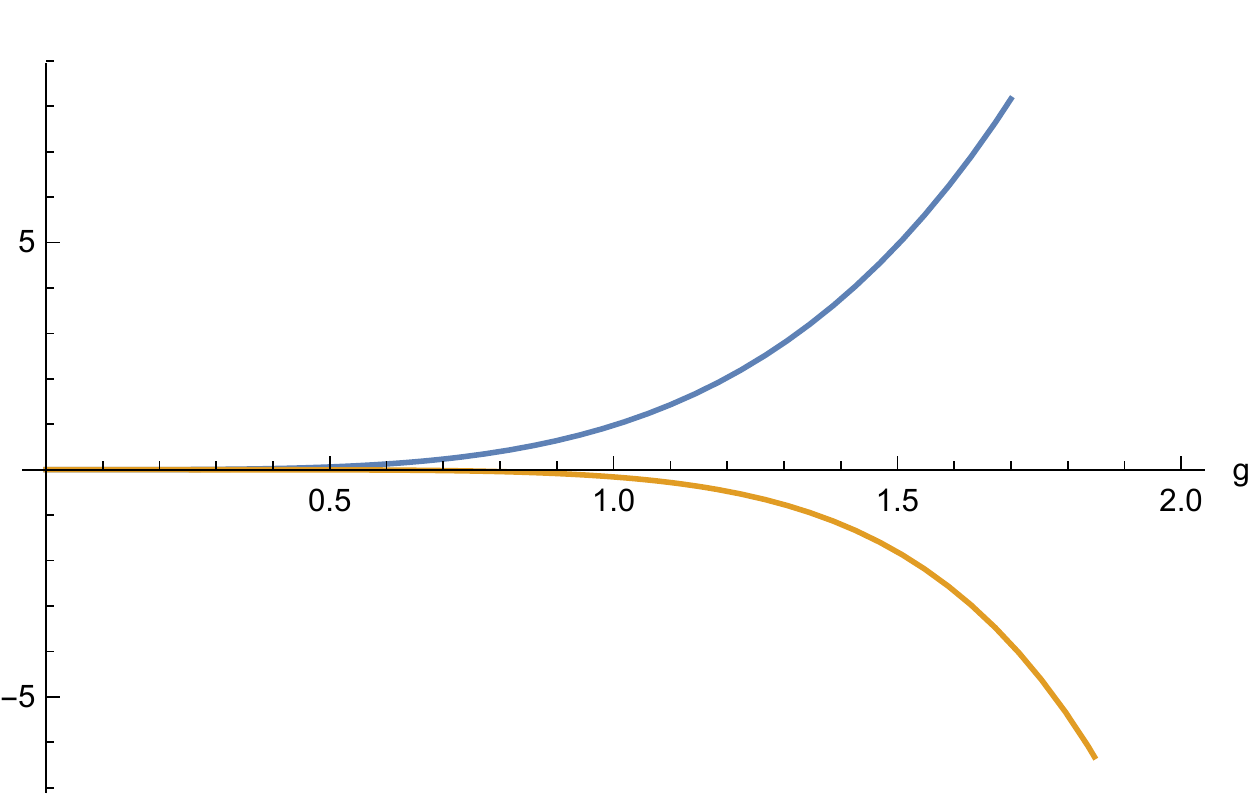}
	\includegraphics[width=0.5\textwidth]{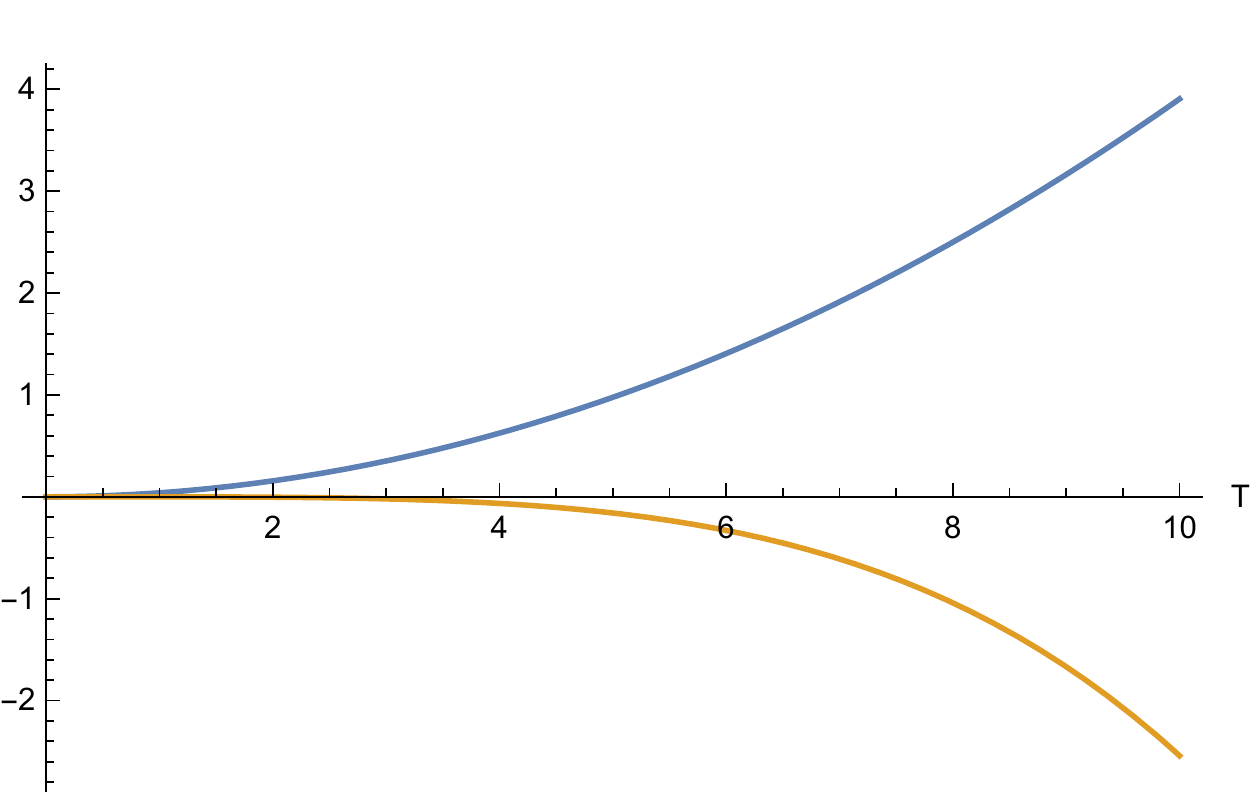}
	\caption{The  the condensate field and the vacuum energy   in one-loop order   for $SU(2)$ as function of the coupling $g$ for $T=10$ (left panel) and as function of the temperature for $g=1$ (right panel).
	}
	\label{fig:4}\end{figure}

In one-loop order, the energy \Ref{3.6} has a non-trivial minimum resulting from the term proportional to $b^{3/2}T$. This picture is not spoiled from the second loop by its $T^3$-term. In one-loop order, the condensate and the effective potential in its minimum are
\eq{3.7}{ b_{min}^{one} &= \frac{9 {a_1}^2 g^4 T^2}{16 \pi ^2},
	&W^{SU(2), \,one}_{min}&=-\frac{\pi ^2 T^4}{15}	-\frac{27 a_1^4 g^6 T^4}{512 \pi ^4}.
}
Again, the first term of the energy is the gluon blackbody radiation. In this approximation, the condensate is always positive, i.e., always  present, and the energy in the minimum is always negative. Plots are shown in Figure \ref{fig:4}.

\begin{figure}[h]
	\includegraphics[width=0.5\textwidth]{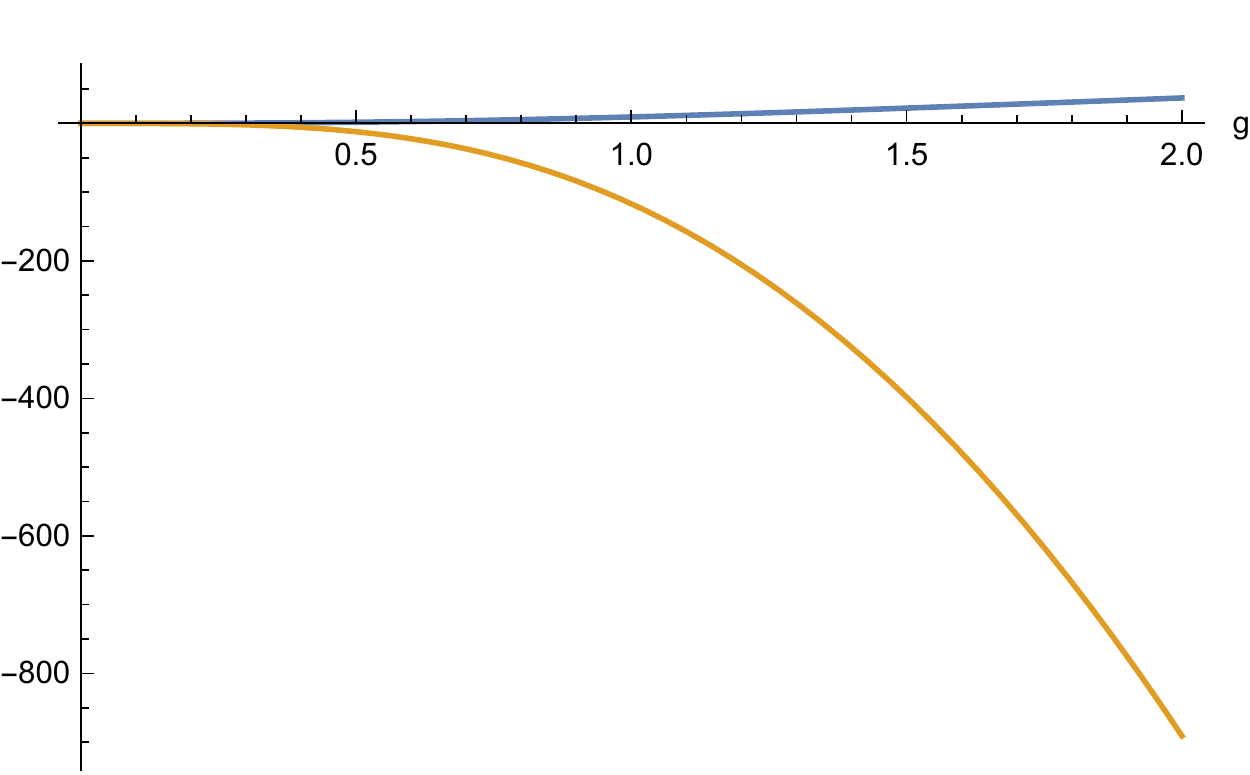}
	\includegraphics[width=0.5\textwidth]{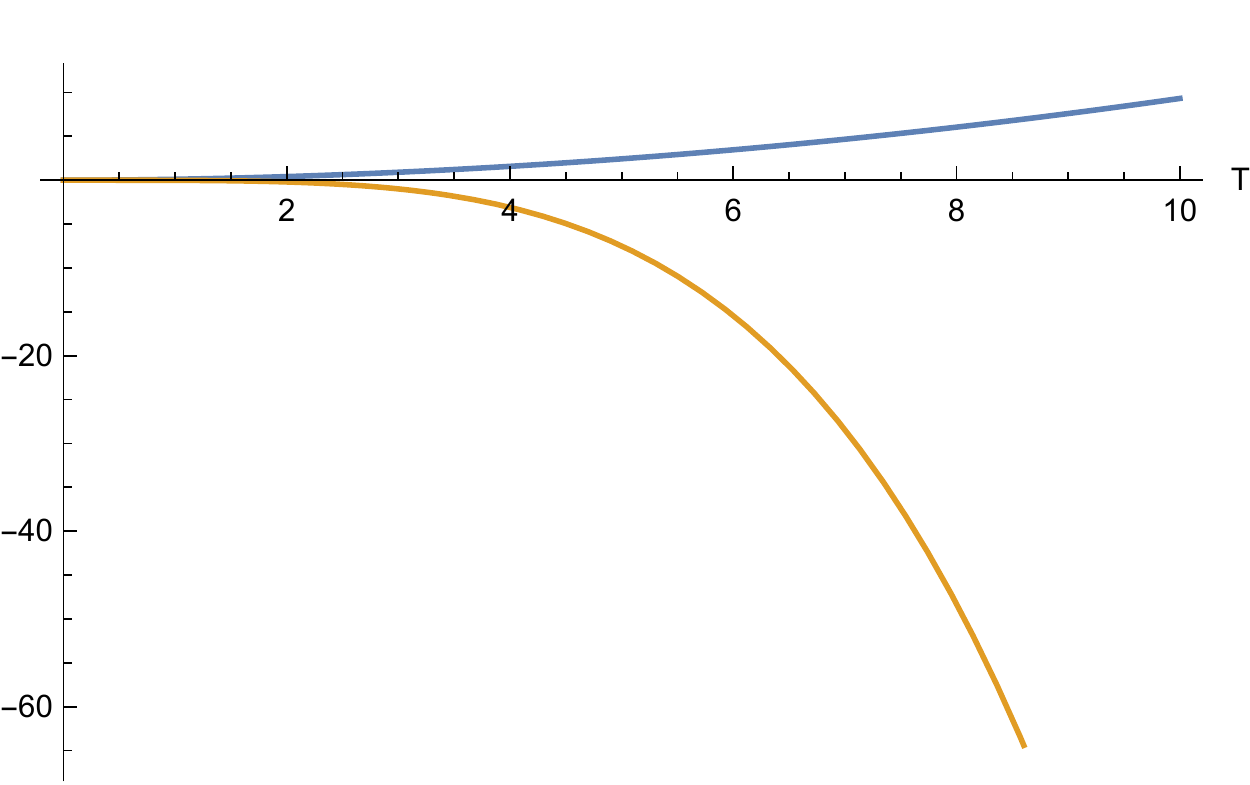}
	\caption{The  the condensate field and the vacuum energy   in two-loop order   for $SU(2)$ as function of the coupling $g$ for $T=10$ (left panel) and as function of the temperature for $g=1$ (right panel).
	}
	\label{fig:5}\end{figure}

In two-loop order, one has to find a root of a third-order polynomial and explicit formulas become quite complicated. Instead, we show the condensate and the effective potential in the minimum, graphically, see Figure \ref{fig:4}. The behavior is similar to the one-loop behavior given by the formulas in eq. \Ref{3.7} and shown in Figure \ref{fig:5}. However, the condensate and the energy grow faster with increasing $g$ or $T$.

\begin{figure}[h]
 	\includegraphics[width=0.5\textwidth]{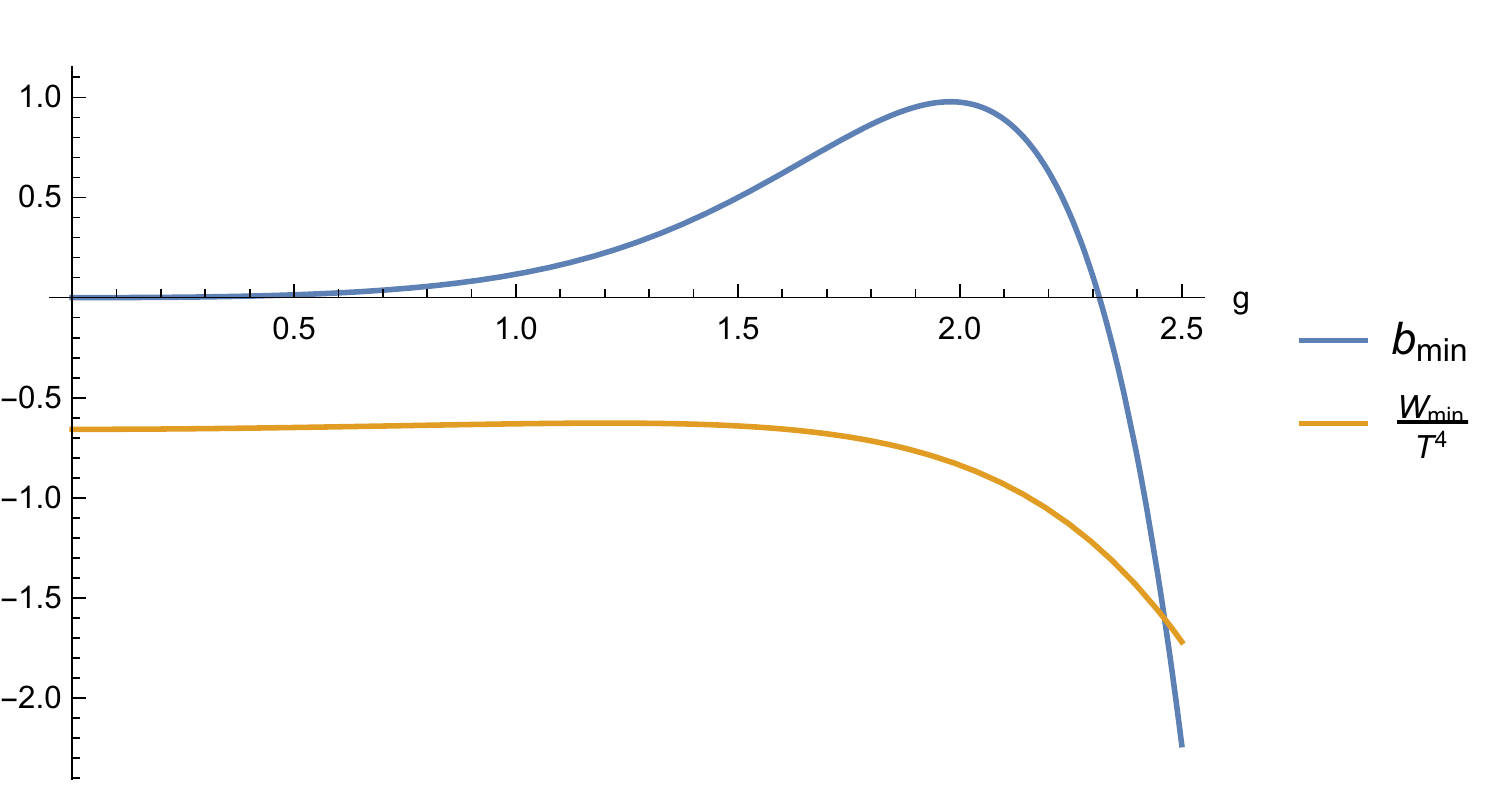}
 	\includegraphics[width=0.5\textwidth]{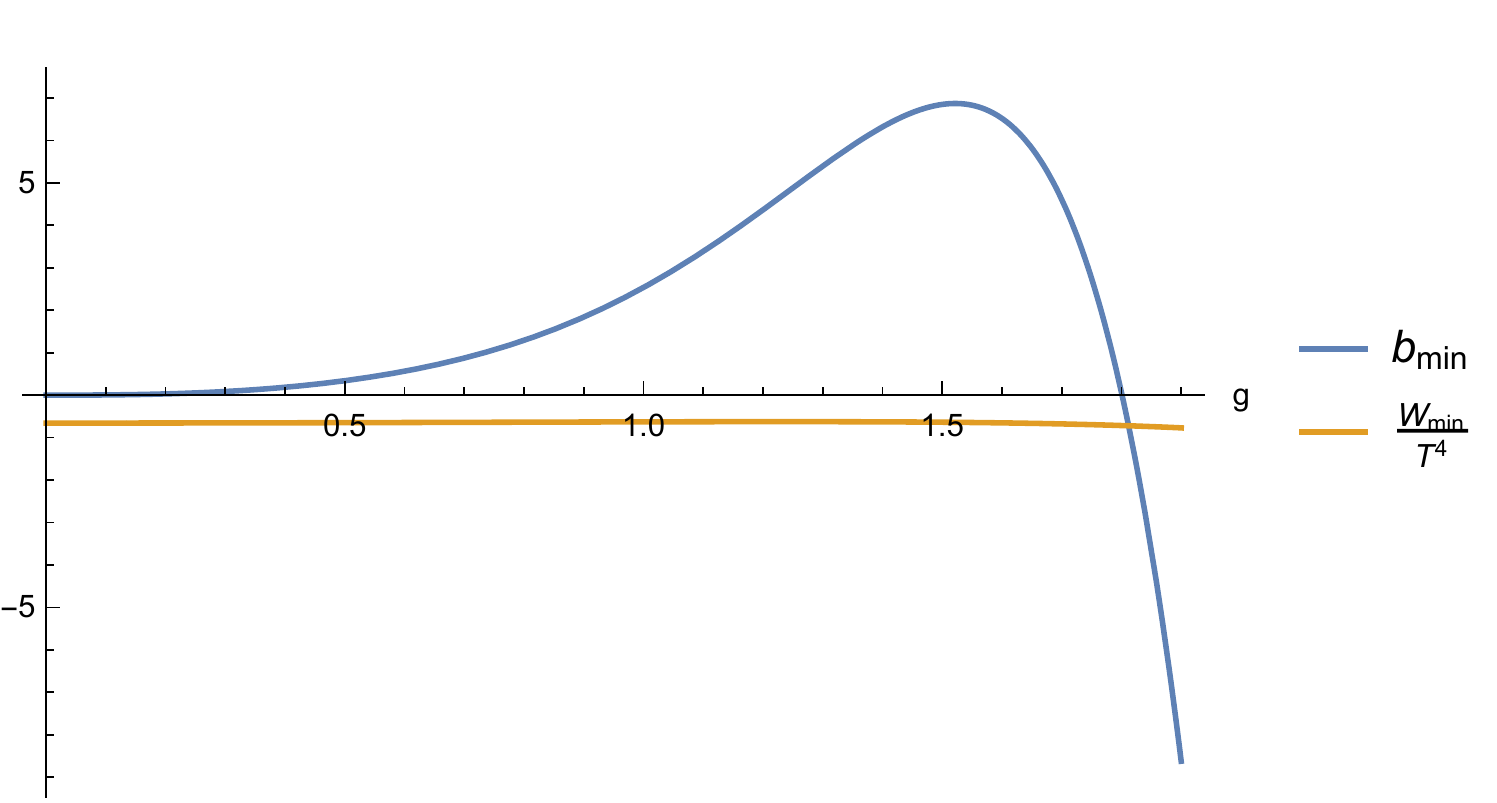}
 	\caption{The   condensate field and the vacuum energy   in two-loop order  from the expansion \Ref{3.8}  as function of the coupling $g$ for $T=1$ (left panel) and   for $T=5$ (right panel).
 	}
	\label{fig:6}\end{figure}

In two-loops, in order to get an expansion in the coupling constant, it is meaningful to take the numerical values in \Ref{3.6}. With these, the expansions read
\eq{3.8}{b_{min} &= 0.0846044 g^{8/3} T^2+0.00575057 g^4 T^2
	+g^{14/3} T^2 (-0.00523858 \ln
	(T)-0.0102511)+O\left(g^5\right),
\\\nn W_{min} &=-\frac{\pi ^2 T^4}{15}+\frac{g^2 T^4}{24}
-0.0107369 g^{10/3} T^4-0.00324209 g^4
T^4
+0.000922241 g^{14/3} T^4
\\\nn	&~~~~~~~~~~~+(0.000332405 \ln
(T)+0.000318908)g^{16/3} T^4 +O\left(g^6\right).
}
As compared with \Ref{3.7}, the powers of $g$ changed; for $b_{min}$ even in leading order and the magnetic field becomes stronger. This perturbative expansion breaks down for sufficiently large $g$, as can be seen from the terms with $\ln T$, which enter with 'wrong' sign. Also this can be seen in Figure \ref{fig:6}, where eq. \Ref{3.8} is plotted. A conclusion is that the coupling $g$ should not much exceed unity.

\section{\label{T4}The minimum of the effective action in the pure $A_0$ case ($b=0$) and comparison with the pure magnetic case}
In this section, we remind the known results for the case of a pure $A_0$-background. We follow the recent paper \cite{bord21-81-998}, where this case was investigated in detail for $SU(3)$. In our case of $SU(2)$ the formulas are even easier. For $b=0$, the effective action \Ref{W2} with \Ref{1.1} is expressed in terms of Bernoulli polynomials. We restrict ourselves to the main topological sector and there to $0\le a\le 1/2$. Here, the effective potential has a minimum at $a= a_{min}$ (see also eq. (6) in \cite{skal92-7-2895}) and takes in this minimum the value $W_{|{a=a_{min}}}=W_{min}$ with
\eq{4.1}{a_{min} &= \frac{3-\xi}{16\pi^2}g^2,
	& W_{min} &= -\frac{\pi^2T^4}{15}+\frac{T^4}{24}g^2
	-\frac{(3-\xi)^2T^4}{192\pi^2}g^4.
}
As mentioned in \cite{bord21-81-998}, \Ref{4.1} coincides with the gauge-invariant result for $\xi=-1$, what we assume in the following. The first term of the effective potential is the gluon black body radiation.

\begin{figure}[h]
	\includegraphics[width=0.5\textwidth]{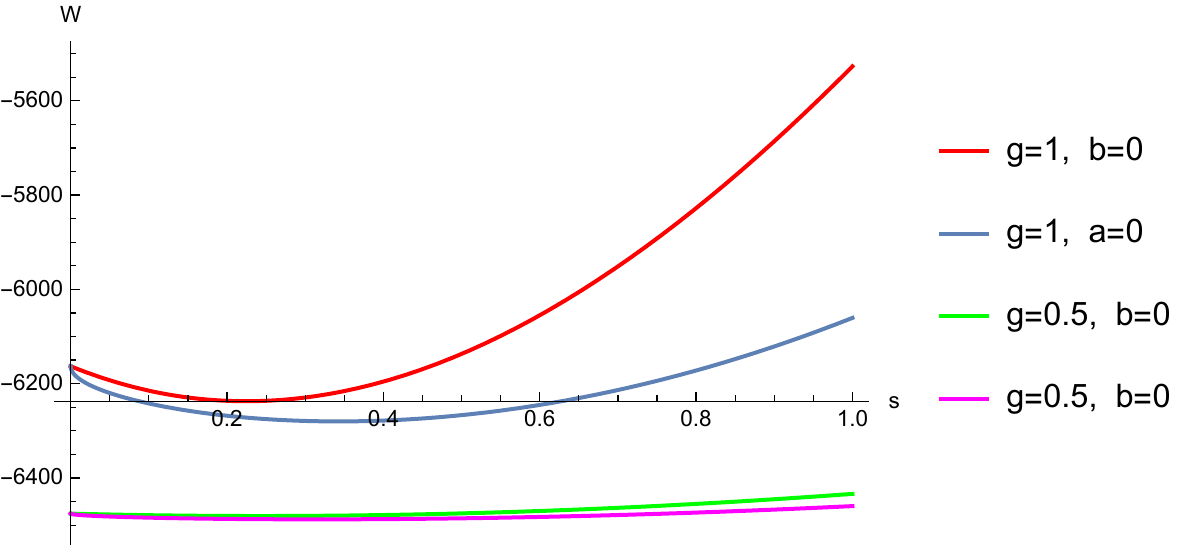}
	\includegraphics[width=0.5\textwidth]{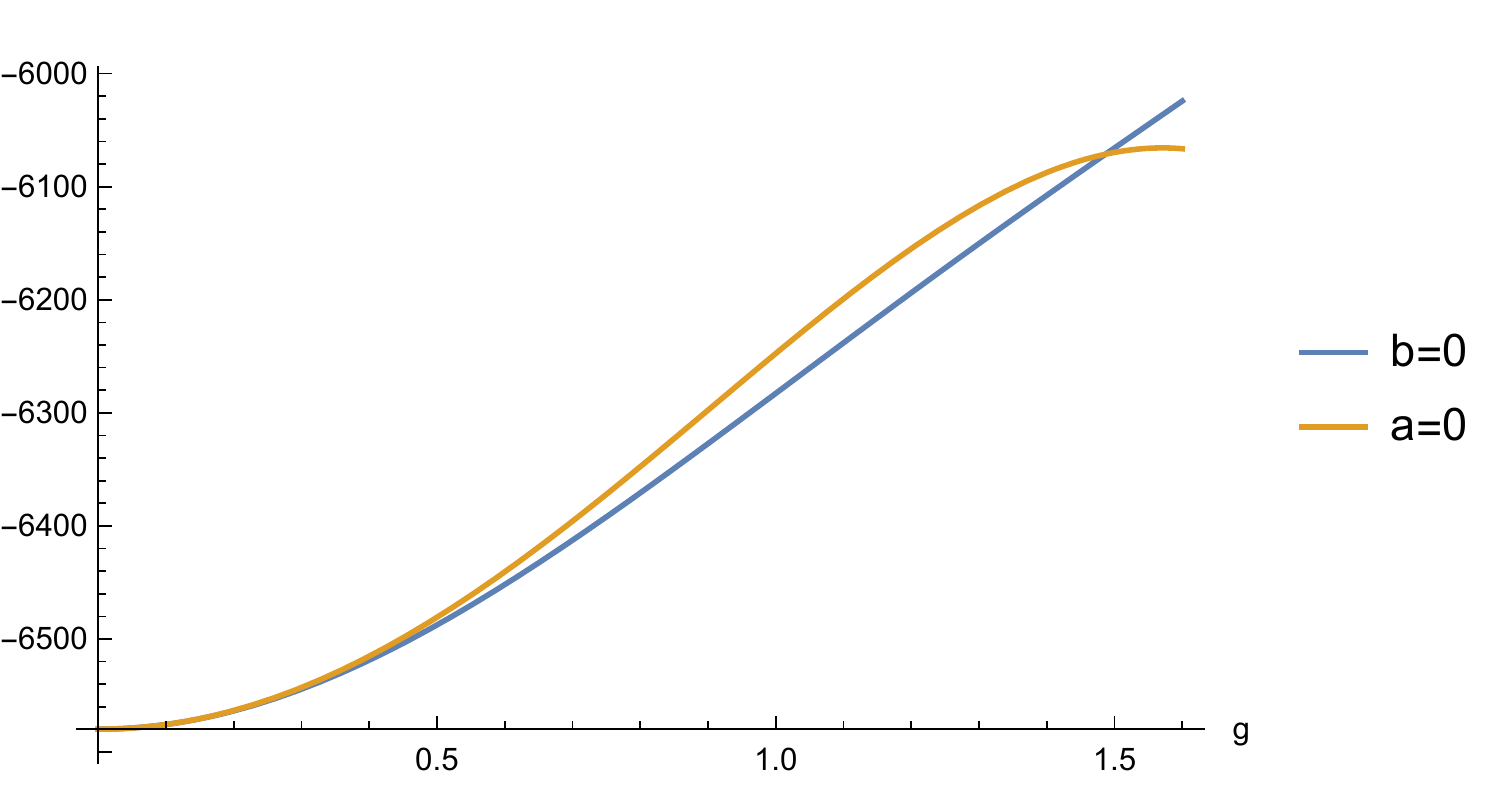}
	\caption{The effective potentials \Ref{3.8} and \Ref{4.1} as functions of the re-scaled variable $a=4 a_{min}s$ and $b=4b_{min}s$  for $T=10$ (left panel). The  minima of the effective potential at $a=0$, \Ref{3.8}, and at $b=0$. \Ref{4.1},  as function of the coupling $g$ for $T=10$   (right panel).
	}
\label{fig:7}\end{figure}

If comparing \Ref{4.1} with the minimal effective potential \Ref{3.8} in the pure magnetic case, it can be seen that in order $g^2$ these coincide. In  Figure \ref{fig:7} (left panel), we show  these effective potentials as functions of the re scaled variables $a=4 a_{min}s$ and $b=4b_{min}s$ with $0<s<1$ (in order to fit into one figure), for two values of the coupling.
The difference between them is
\eq{4.2}{W_{min}^{\Ref{3.8}}-W_{min}^{\Ref{4.1}}
			 =-0.0107369 g^{10/3} T^4-0.00324209 g^4
			 T^4+\dots\,.
}
These two minima are shown in Figure \ref{fig:7} (right panel). The difference between them is of order higher than $g^2$.

\section{\label{T5}The numerical evaluation of the effective potential as function of both, $a$ and $b$}
\begin{figure}[h]
	\includegraphics[width=0.5\textwidth]{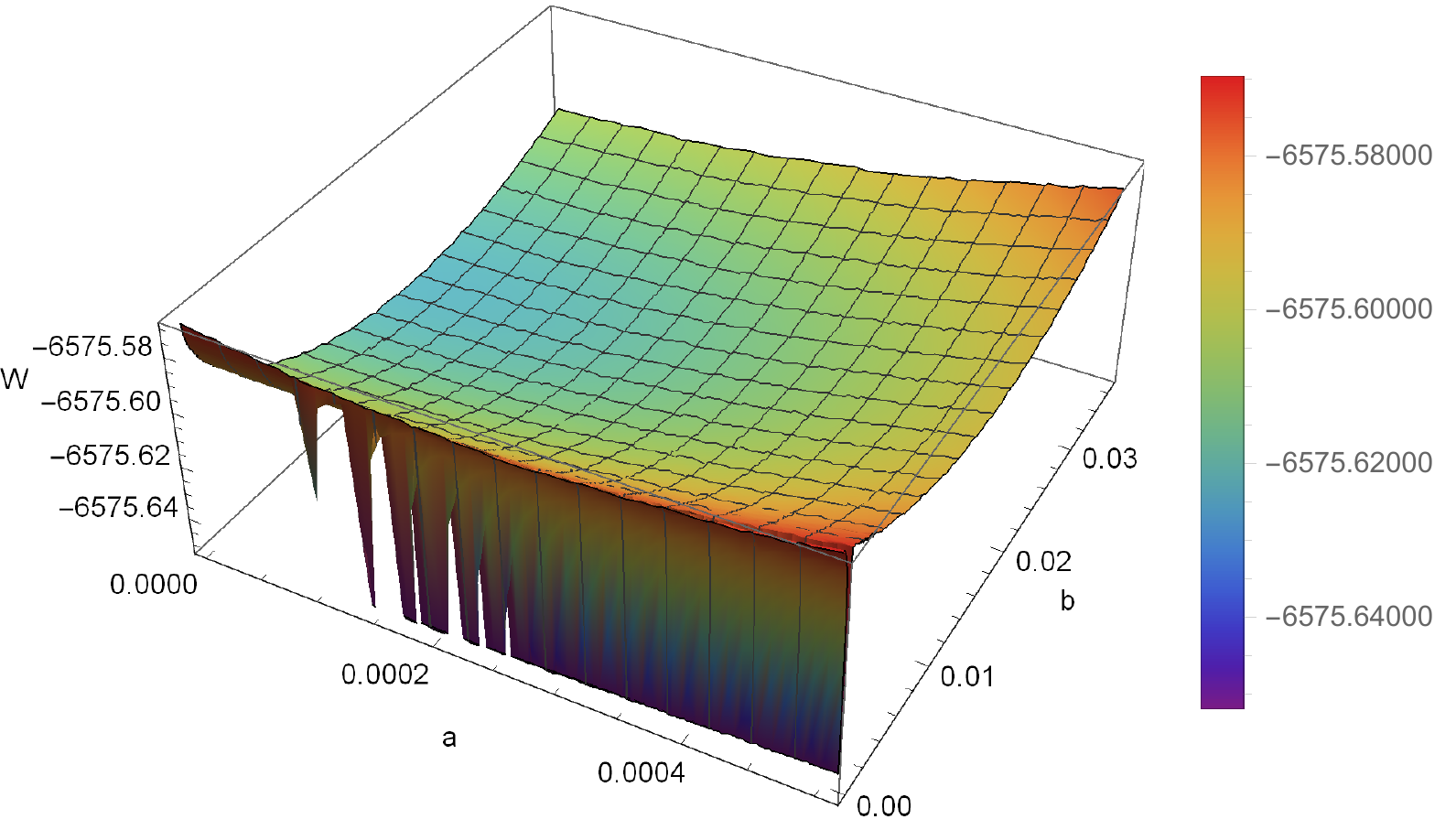}
	\includegraphics[width=0.5\textwidth]{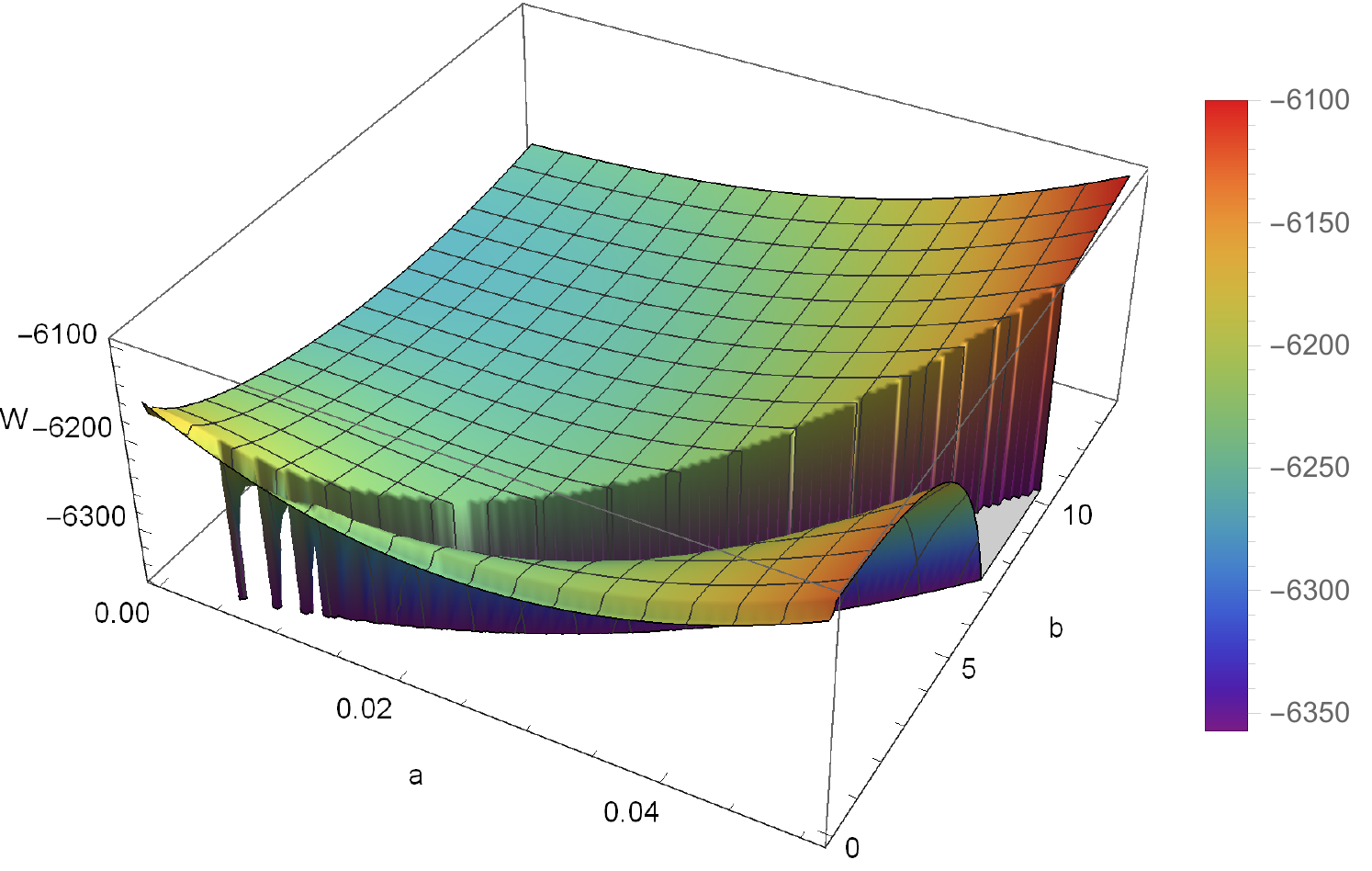}
	\caption{The real part of the effective potential $W_{gl}^{SU(2)}$, \Ref{W2a}, as function of the $A_0\frac{2\pi T}{g}$ and the magnetic background $H^3=\frac{b}{g}$ (see eq. \Ref{ab}) for $T=10$ (in arbitrary units) and $g=0.1$ (left panel) and $g=1$ (right panel).
	}
	\label{fig:8}\end{figure}

\begin{figure}[h]
	\includegraphics[width=0.5\textwidth]{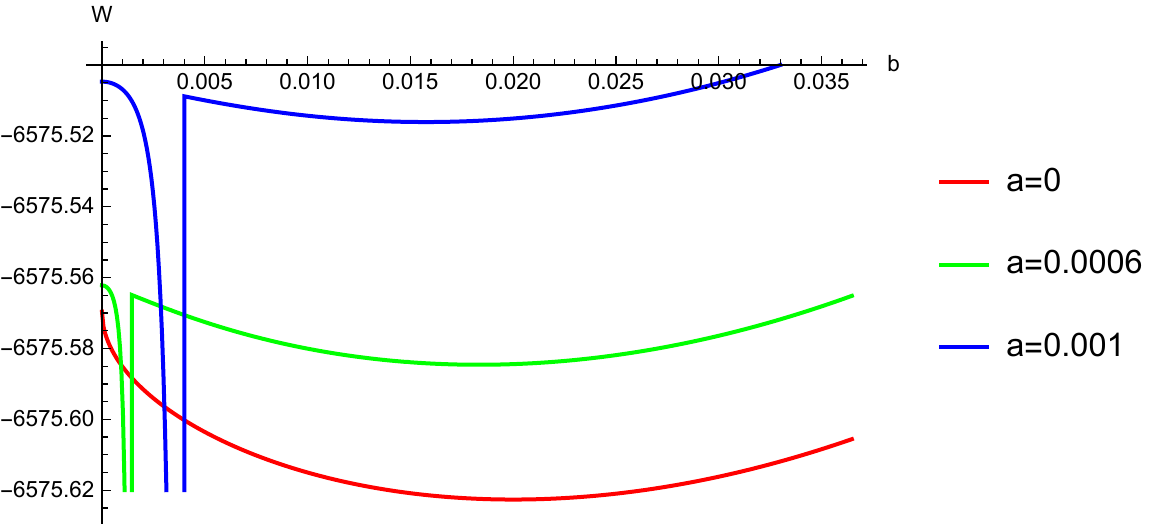}
	\includegraphics[width=0.5\textwidth]{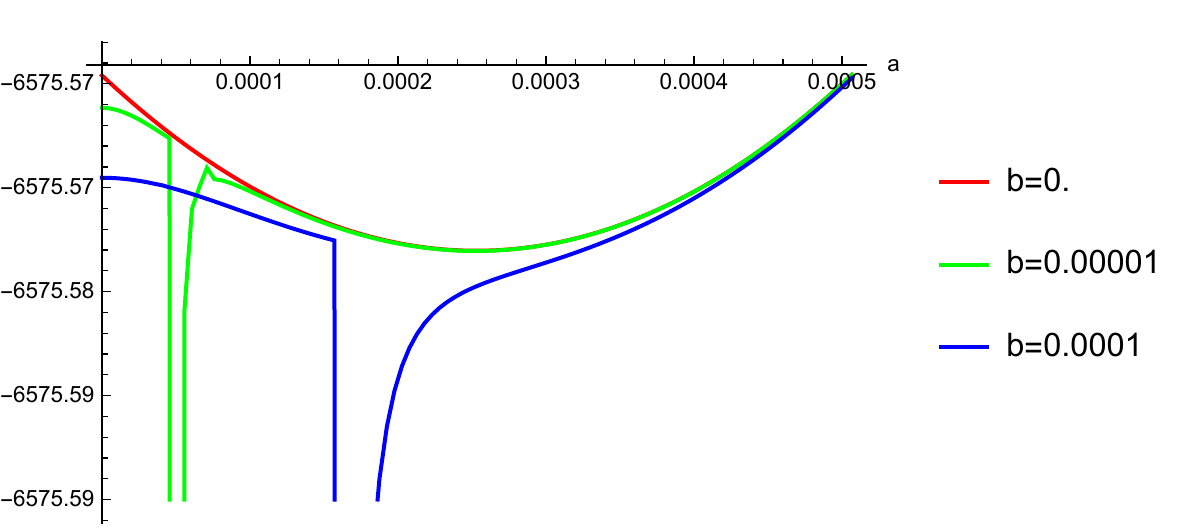}
	\caption{Sections of   Figure \ref{fig:8} (left panel) near the axes.	}
	\label{fig:9}\end{figure}

The formulas of Section \ref{T2} allow for an numerical evaluation of the effective potential \Ref{W2a} as a function of two parameters, $a$ and $b$, \Ref{ab}. We remind, that we put the parameter $\mu$, which measures all dimensional quantities, equal to unity, $\mu=1$. Parameters are the temperature $T$ and the coupling $g$. We demonstrate the results in Figure \ref{fig:8} for two different couplings.
In both panels, the ranges of $a$ and $b$  are chosen to include the minima on each axis.
Details near the axes are shown in Figure \ref{fig:9}.
It is seen, that a deep minimum appears immediately when one leaves the axis, i.e., for small $b$ in the left panel and for small $a$ in the right panel. These minima are seen in  Figure \ref{fig:8} as a valley. For smaller coupling $g$, these are narrower than for larger $g$.

The origin of these minima can be traced back to eqs. \Ref{2.19}, \Ref{2.21}  and \Ref{2.22}, showing singular behavior. Also, the imaginary parts \Ref{2.16} show such behavior. Clearly, these minima are a consequence of the imaginary part present in the given approximation, i.e., with two loops. As mentioned in eq. \Ref{2.17}, for small $b$ ($b<(2\pi T a)^2$), there is no imaginary part and the parabola $b=(2\pi T)^2$ in the $(a,b)$-plane  marks the onset of the imaginary part.

\section{\label{T6}Conclusions}
In the forgoing sections, we investigated in detail the effective potential in $SU(2)$ gluodynamics in the background of both, $A_0$  and magnetic field $H^3$. The basic results are shown in  Figure \ref{fig:8}, using the notations \Ref{ab}. We confirmed the known results for, separately, $A_0=0$ and $H^3=0$,i.e.,  for the minima on the axes.

As can be seen from  Figure \ref{fig:8}, and in detail from  Figure \ref{fig:9},  as soon as one goes from $A_0=0$ into the $(A_0,H^3)$-plane, the imaginary part known from the Savvidy vacuum sets in. The leading contribution follows from the product of $B_1^{ta}(a,b)$ and $B_3^{ta}(a,b)$, \Ref{2.22}, for $\ell=0$,  in $W_{gl}^{SU(2)}$, \Ref{2.19}, to be
\eq{6.1}{W_{gl}^{SU(2)} &\sim
	-\frac{a^2b^2T^4}{4((2\pi T a)^2-b)^2}
}
for $b \lesssim (2\pi T a)^2$. Beyond, this contribution disappears and the imaginary part from \Ref{2.16} sets in. Thus,  we have to interpret the 'valley' as a remnant of the imaginary part. We did not show any picture of the imaginary part since it is not interesting beyond its mere existence.

As mentioned in the Introduction, there is a claim in \cite{skal00-576-430} that summation of the ring diagrams, which is beyond what is used in the present paper, removes the imaginary part while keeping the minimum of the real part. Indeed, with eq. \Ref{3.6} we reproduce, up to differences in numerical coefficients, eq. (37) in \cite{skal00-576-430} up to the ring contributions.

As for the minimum of the effective action as a function of the two parameters $a$ and $b$, one can see from the  Figure \ref{fig:8} that it is clearly dominated by the 'valley' around the onset of the imaginary part. As for the minima on the axes, in leading order in the coupling $g$ these are of equal depth, including higher orders the magnetic minimum becomes deeper, as can be seen from eq. \Ref{4.2} and  Figure \ref{fig:7}. This picture is also confirmed by  Figures \ref{fig:8} and \ref{fig:9}.
When starting from $b=0$, with increasing $b$ the $A_0$-minimum gets lifted as can be seen clearly on  Figure \ref{fig:9}, left panel.
When starting from $a=0$, with increasing $a$, the Savvidy minimum becomes deeper. However, in both cases, the picture is dominated by the mentioned above remnant of the imaginary part.

From the above discussions, we come to the conclusion that at the considered two-loop level,  no conclusive judgment on the minimum of the effective potential is possible. It is clear that one has  to go beyond and  do the summation of ring diagrams with nonzero  $A_0$, at least. In this connection, it is worth mentioning the lattice result \cite{demc08-41-165051}, where a minimum in the ($A_0,b$)-plane was found. As a lattice result, it goes, of course, beyond any perturbative re-summation and cannot be compared with the current calculation. {However, it may serve as a motivation to go further.

We conclude with a remark on the high-T approximation discussed in Section \ref{T2.2}. In this paper, we took $T=10$ for the numerical examples. Several checks showed that
the numbers produced this way are very close to the corresponding numbers produced with the asymptotic expansions  as long as  either $a=0$ or $a\gtrsim 0.1$, down to $T^2/b\sim 1$. For small, but non-zero values of $a$, the high-T asymptotic cannot be used. Especially, the singular behavior seen in eq. \Ref{6.1}, is completely missing in  the high-T expansion although the tachyonic mode is included there.

\end{document}